\documentclass[manuscript]{acmart}

\usepackage{tikz}
\usepackage{amsmath}
\usepackage{lipsum}

\usepackage{filecontents}
\usepackage{balance}
\usepackage{adjustbox}
\usepackage{algorithm}
\usepackage{amsmath}
\usepackage[noend]{algpseudocode}
\usepackage{xcolor}
\usepackage{url}
\usepackage{soul}

\usepackage{pifont}

\newcommand{\cmark}{\ding{51}}
\newcommand{\xmark}{\ding{55}}

\usepackage{subfigure}
\usepackage{balance}
\usepackage{caption}

\usepackage{colortbl}

\usepackage[utf8]{inputenc}
\usepackage[english]{babel}

\def\BibTeX{{\rm B\kern-.05em{\sc i\kern-.025em b}\kern-.08em
		T\kern-.1667em\lower.7ex\hbox{E}\kern-.125emX}}

\newcommand{\ie}{{\em i.e., }}
\newcommand{\eg}{{\em e.g., }}
\newcommand{\myverb}{\fontsize{10}{48}\usefont{OT1}{lmtt}{b}{n}\noindent }

\usepackage{tcolorbox} \newcommand{\ciao}[1]{{\setlength\fboxrule{0pt}\fbox{\tcbox[colframe=black,colback=white,shrink tight,boxrule=0.5pt,extrude by=1mm]{\small #1}}}}

\AtBeginDocument{%
  \providecommand\BibTeX{{%
    \normalfont B\kern-0.5em{\scshape i\kern-0.25em b}\kern-0.8em\TeX}}}

\begin{document}
\title{A Survey on DNS Encryption: Current Development, Malware Misuse, and Inference Techniques}

\author{Minzhao Lyu}
\affiliation{
  \institution{University of New South Wales}
  \city{Sydney}
  \country{Australia}
}
\email{minzhao.lyu@unsw.edu.au}

\author{Hassan Habibi Gharakheili}
\affiliation{
	\institution{University of New South Wales}
	\city{Sydney}
	\country{Australia}
}
\email{h.habibi@unsw.edu.au}

\author{Vijay Sivaraman}
\affiliation{
	\institution{University of New South Wales}
	\city{Sydney}
	\country{Australia}
}
\email{vijay@unsw.edu.au}
\renewcommand{\shortauthors}{Lyu, et al.}

\begin{abstract}
	
The domain name system (DNS) that maps alphabetic names to numeric Internet Protocol (IP) addresses plays a foundational role in Internet communications.
By default, DNS queries and responses are exchanged in unencrypted plaintext, and hence, can be read and/or hijacked by third parties. To protect user privacy, the networking community has proposed standard encryption technologies such as DNS over TLS (DoT), DNS over HTTPS (DoH), and DNS over QUIC (DoQ) for DNS communications, enabling clients to perform secure and private domain name lookups.
We survey the DNS encryption literature published from 2016 to 2021, focusing on its current landscape and how it is misused by malware, and highlighting the existing techniques developed to make inferences from encrypted DNS traffic.
First, we provide an overview of various standards developed in the space of DNS encryption and their adoption status, performance, benefits, and security issues.
Second, we highlight ways that various malware families can exploit DNS encryption to their advantage for botnet communications and/or data exfiltration.
Third, we discuss existing inference methods for profiling normal patterns and/or detecting malicious encrypted DNS traffic.
Several directions are presented to motivate future research in enhancing the performance and security of DNS encryption.

\end{abstract}


\setcopyright{acmcopyright}
\acmJournal{CSUR}
\acmYear{2022} \acmVolume{1} \acmNumber{1} \acmArticle{1} \acmMonth{1} \acmPrice{15.00}\acmDOI{10.1145/3547331}


\maketitle

\section{Introduction}\label{sec:Intro}

\begin{figure}[!t]
	\begin{center}
		\includegraphics[width=0.97\textwidth]{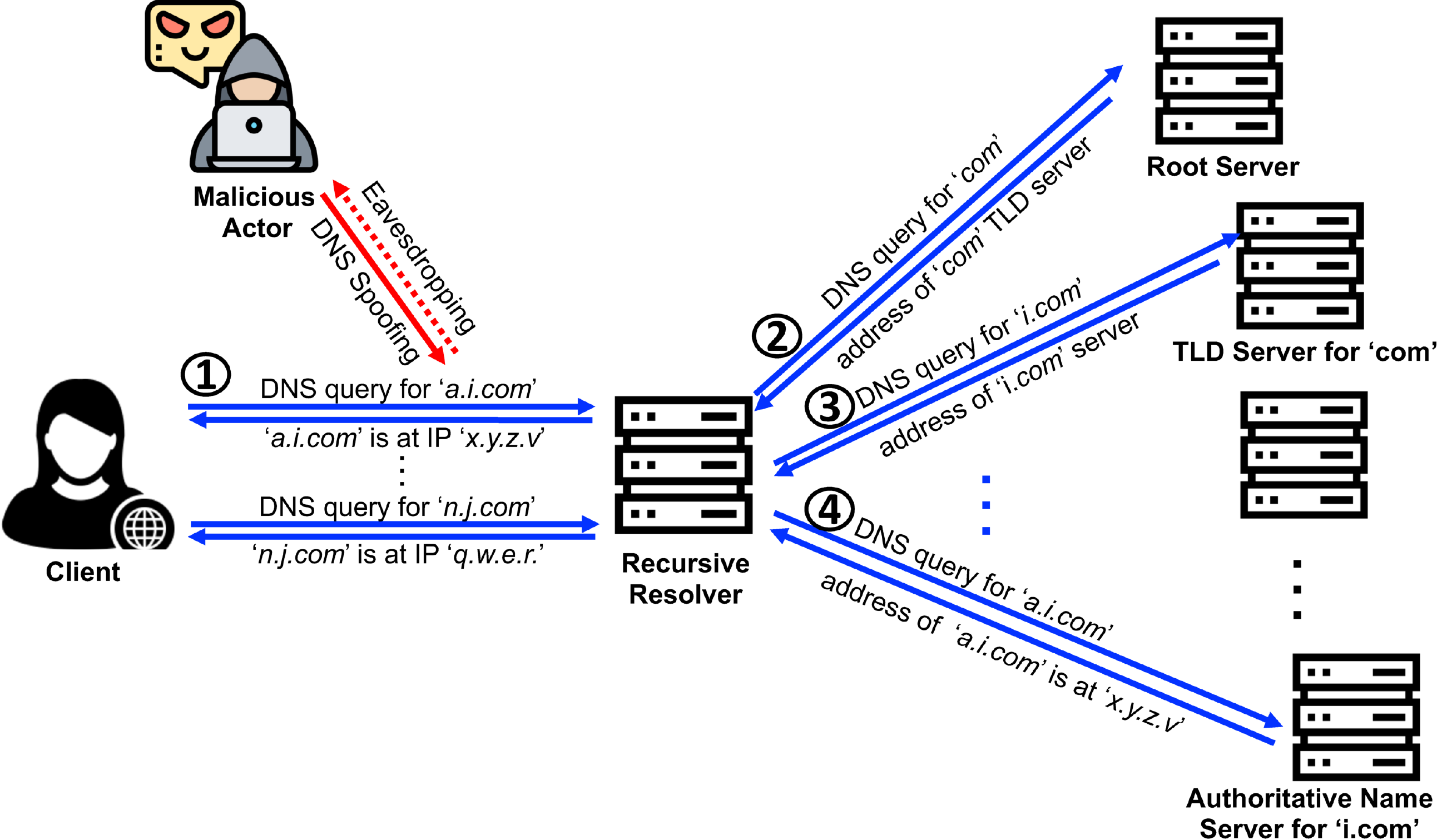}
		\caption{A visual example of DNS lookup activities from a client and the potential threats introduced by malicious actors.}
		\label{fig:DNSLookup}
	\end{center}
\end{figure}

The domain name system (DNS) protocol \cite{rfcDNS1987} takes the responsibility of converting human-readable domain names into machine-friendly IP addresses and vice versa, which is critical to the Internet communications today.
We start by illustrating the basic operation of DNS in Fig.~\ref{fig:DNSLookup}:
a client embeds its question (typically for domain names) in DNS queries and sends them to DNS resolvers for lookups of the answers (step \ciao{1}). The resolver then performs recursive lookups of the questioned domain name by successively/recursively querying a root server, top-level-domain (TLD) server, and authoritative name server (steps \ciao{2}, \ciao{3}, and \ciao{4} in Fig.~\ref{fig:DNSLookup}, respectively).

The DNS ecosystem presents various types of security and privacy risks that could be exploited by malicious actors.
For example, DNS resolvers and name servers often become attractive targets of denial-of-service (DoS) attacks that slow down or even paralyze their services. For countermeasures, there exist specialized security systems and appliances that are able to detect and mitigate DoS attacks on DNS infrastructures included \cite{BHiteshCCS2008,MLyuTNSM2021}.
It is also a common practice for attackers to mislead clients by fake responses (DNS spoofing and/or hijacking), directing them to malicious servers. To tackle this problem, DNSSEC \cite{rfcDNSSEC2005} was proposed for protecting the data integrity of DNS. It provides cryptographic verification through digital signatures that can be used to validate the records delivered in a DNS response from the authoritative DNS server.
Furthermore, as shown in Fig.~\ref{fig:DNSLookup}, DNS queries and responses are communicated in plaintext via UDP  transport-layer protocol which offers various benefits such as low computational overheads, fast resolution, and ease of deployment and management \cite{DoHThreatLandscape}. However, malicious actors take advantage of unencrypted contents, putting user privacy at risk \cite{YORK201041} or spoofing \cite{XBaiWISM2011} DNS lookups between clients and resolvers.
To address privacy concerns, encrypting DNS lookups (exchanged between client and resolver) has been developed and promoted, which is the focus of this survey.

Various techniques of DNS encryption have been developed, experimented with, and deployed across the Internet over the past few years to protect user privacy.
Some early proposals such as DNSCurve \cite{rfcDNSCurve2010} and DNSCrypt \cite{DNSCrypt2018} were introduced in 2008 and 2011 \cite{MSehring2020}, respectively. However, none of them became officially standardized, and hence are not widely deployed by the Internet community.
To facilitate the adoption of DNS encryption, the Internet Engineering Task Force (IETF) has proposed a series of standards (RFCs) starting from 2016, relying upon existing secure/private protocols, including DNS over TLS (DoT) \cite{rfcDNSTLS2016}, DNS over HTTPS (DoH) \cite{rfcDNSHTTPS2018}, and DNS over QUIC (DoQ) \cite{rfcDNSDoQ2021}.
Currently, the public adoption of encrypted DNS (compared with plaintext DNS) is still relatively low, but several major Internet technology and service providers such as Google, Cloudflare, Cisco, and Alibaba have launched their encrypted DNS resolvers \cite{DOHServerLists}, fueling the growth of encrypted DNS traffic on the Internet \cite{SGarciaARXIV2021}.

Regardless of the benefits brought by DNS encryption, some practical issues and security problems have been identified by the existing literature, impacting its public adoption. For example, encrypted resolution may not be feasible if encryption-enabled DNS resolvers fail to provide a valid certificate.
Unfortunately, as reported in \cite{CLuIMC2019}, lack of best practice encryption setups is not uncommon in today's public encrypted DNS resolvers. Moreover, inappropriately configured DNS resolvers are vulnerable to fallback (downgrade) attacks (\eg SSL stripping), forcing resolvers to return  DNS responses in plaintext.

\begin{figure*}[!t]
	\begin{center}
		\includegraphics[width=1\textwidth]{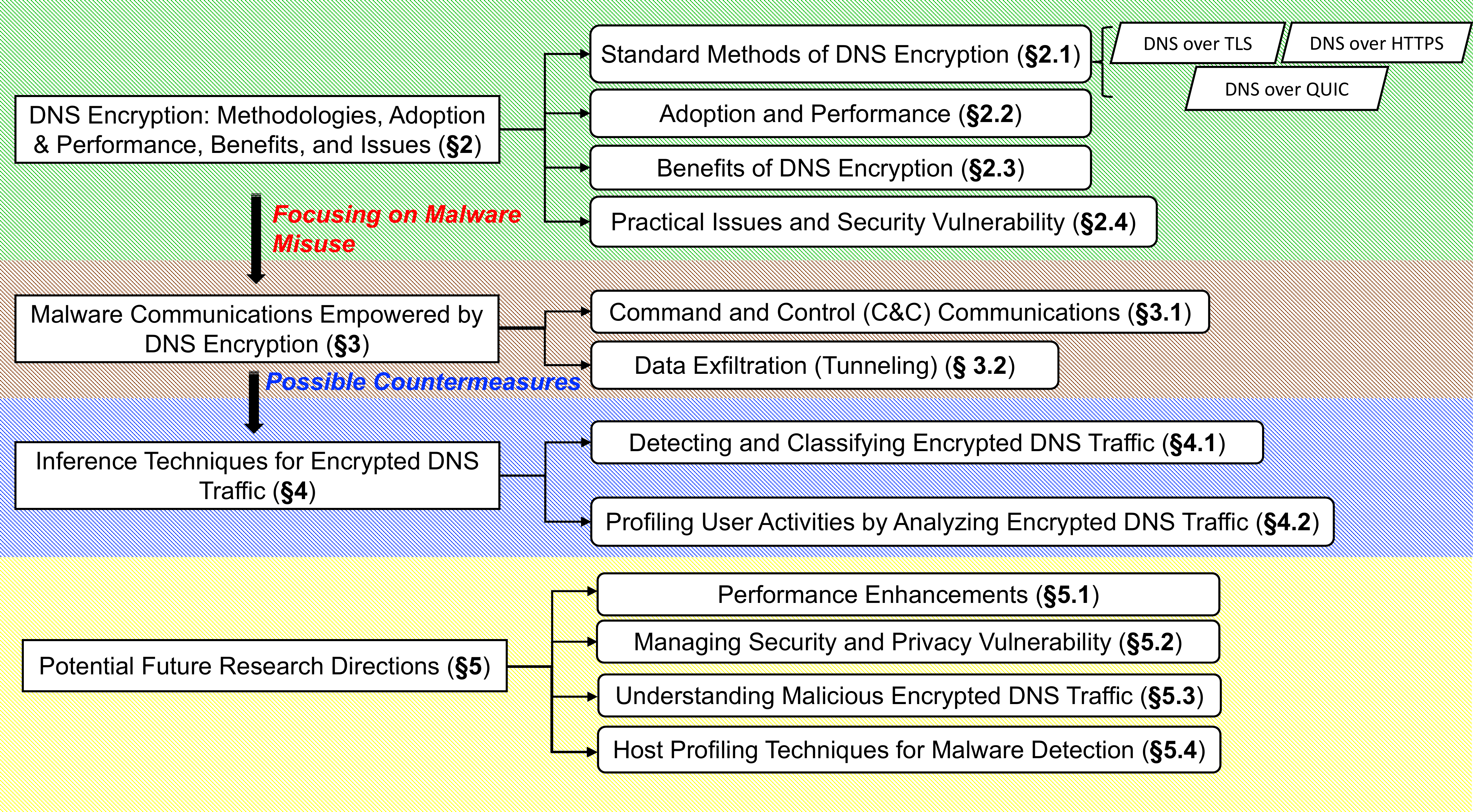}
		\caption{Key contributions and structure  of this survey paper.}
		\label{fig:keyLogistic}
	\end{center}
\end{figure*}

Among those practical issues, the misuse of DNS encryption by malware is one of the problems that has attracted attention from the security community. Cybercriminals can easily hide their identities and activities via encrypted DNS traffic to bypass legacy security tools and appliances. Many monitoring and detection methods today rely on DNS packet content inspection, which is highly effective in processing plaintext-based DNS communications. Existing literature has reported that malicious actors use encrypted DNS for command-and-control (C\&C) communications and/or exfiltrating and tunneling data between malware-infected devices and cloud-based servers -- further details will be discussed later in \S\ref{sec:SurveyMalware}.
To tackle the malicious usage of DNS encryption by malware, researchers have developed various analytical methods to detect encrypted DNS packets from network traffic and classify benign and anomalous streams. In addition, there are also research works that profile user behaviors by analyzing encrypted DNS traffic. Prior works collectively provide solid insights into challenges and opportunities, and motivate future researches in detecting and fingerprinting malware-infected devices that utilize encrypted DNS for stealthy communications.

This survey is the first to comprehensively review relevant literature (published from 2016 to 2021) on DNS encryption techniques, their opportunities, and risks.
To guarantee that our survey has a full coverage of peer-reviewed papers in the domain of DNS encryption, we searched seven title keywords from eight major scientific digital libraries such as IEEE Xplore and ACM Digital Library. 
Research papers we surveyed are from a wide range of academic journals, conferences, and workshops in general computer networking (\eg IEEE TNSM, Computer Networks), Internet measurement (\eg ACM IMC, PAM), and cybersecurity (\eg Computers \& Security and USENIX Security). 
In addition, we incorporated relevant Internet standard documents (RFCs), technical reports from reputable organizations, and research papers that do not directly focus on encrypted DNS to cover certain key points around DNS encryption.

\begin{table*}[!t]
	\caption{List of references cited in each category of DNS encryption research.}
	\centering
	\normalsize
	\renewcommand{\arraystretch}{1.2}
	\begin{tabular}{|p{0.2cm}|p{7.6cm}|p{6cm}|}
		\hline
		\rowcolor[rgb]{ .906,  .902,  .902}  &	\textbf{Category of topics covered by this survey}               & \textbf{List of references} \\ \hline
		\S\ref{sec:SurveyDevelopment}&	Standardized DNS encryption methodologies & \cite{PWu2019,rfcDNSTLS2016,rfcTLS2008,RKartch2017,HTTPSOnly,rfcDNSHTTPS2018,DFisher2020,rfcDNSDoQ2021,rfcQUIC2021,AGhedini2019,TBucutiISI2015,GCarlucciSAC2015} \\ \hline
		\S\ref{sec:SurveyDevelopment}&	Adoption and performance       & \cite{MVale2019,DNSEncryptedResolver,DoHResolver,VBagirov2020,SGarciaARXIV2021,MCWasastjernaECJ2018,TANideck2019,CHesselmanIC2020,PWu2019,ANisenoffDNSPW2021,SDeckelmann2020,ZYanFGCS2020,HShulmanWPES2014,AHounselANRW2019,TBottgerIMC2019,KBorgolteEJ2019,AHounselWWW2020,ESMbeweAFRICOMM2021,TNDoanPAM2021,AHounselPAM2021,RChhabraIMC2021}    \\ \hline
		\S\ref{sec:SurveyDevelopment}&	Benefits of DNS encryption                        & \cite{HShulmanWPES2014,ZYanFGCS2020,KBumanglagICICT2020,MAnagnostopoulosCS2013,LZhuSP2015,NNayakICCSIT2010,PJeitnerDSNS2020}\\ \hline
		\S\ref{sec:SurveyDevelopment}&	Practical issues and security vulnerability               & \cite{LJinWWW2021,SRiveraARES2020,YNakatsukaACSAC2019,YNakatsukaDT2021,CLuIMC2019,jahromi2021comparative,DShinACSAC2011,QHuangFOCI2020,RHouserCoNEXT2019,rfcDNSPadding2018,KHynekIEMCON2020,JBushartFOCI2020,KRadinsky2015,PSchmittANRW2019,rfcODNS2021,AHounselANRW2021,MLyuPAM2019,JAhmedTNSM2020,MLyuTNSM2021,DoHThreatLandscape,PBischoff2021,ZYanFGCS2020,HHabibiGharakheiliCAN2017,CCimpanu2019,hynek2020doh,MSinghDI2019,KAlieyanNCA2017}\\ \hline
		\S\ref{sec:SurveyMalware}&	Malware misuse: C\&C communications & \cite{CommandAndControl,KXuTDSC2013,JLeeWSNP2010,AUdiyonoICONETSI2020,YZengCN2021,LBilgeNDSS2011,CJDietrichEC2ND,CPatsakisCS2020,Godlua2019,PsiXBot2019,KBumanglagICICT2020,MGrillIM2015}\\ \hline
		\S\ref{sec:SurveyMalware}&	Malware misuse: data exfiltration (or tunneling) & \cite{FUllahJNCA2018,ANadlerCS2019,BSabirCS2022,CCimpanu2020,ExfilDoHTool2,ExfilDoHTool,XHuDSN2016,JAhmedTNSM2020,DHaddonICGS32019} \\ \hline
		\S\ref{sec:SurveyDetection}&	Detecting and classifying encrypted DNS traffic              &  \cite{CPatsakisCS2020,DVekshinARES2020,DVekshinARES2020,vekshin_dmitrii_2020_3906526,LCsikorESP2021,CKwan2021FOCI,MMontazeriShatooriDASC2020,CIRACICDoHBrw2020,SKSingh3ICT2020,YMBanadakiJCSA2020,MBehnkeAccess2021}\\ \hline
		\S\ref{sec:SurveyDetection}&	Profiling user activities by analyzing encrypted DNS analysis &\cite{RHouserCoNEXT2019,rfcDNSPadding2018,SDSibyNDSS2020,siby2018dns,MMuhlhauserARES2021,MMulhauser2021,GVarshneyCOMSNETS2021} \\ \hline
	\end{tabular}
	\label{tab:paperCategory}
\end{table*}

\subsection{Contributions}
The contributions of this survey paper can be summarized as follows.
\begin{itemize}
	\item \textbf{First}, we outline the current development of DNS encryption, highlighting aspects like standard techniques (\ie DoT, DoH, and DoQ),  the current status of their adoption across the Internet, performance analysis, benefits, practical issues, and security vulnerabilities identified by the current literature.
	\item \textbf{Second}, we discuss how DNS encryption can be misused by malware for purposes including command-and-control (C\&C) communications and data exfiltration, with highlights on the (in)effectiveness of existing countermeasures originally developed for plaintext DNS when applied to encrypted DNS.
	\item \textbf{Third}, we survey the current analysis and inference methods for detecting encrypted DNS traffic from generic encrypted network streams (\eg HTTPS), classify malicious encrypted DNS communications, and fingerprint host profiles by analyzing encrypted DNS traffic. They provide strong references and motivations for future research in detecting malware-infected devices that exploit encrypted DNS protocol.
	\item \textbf{Last}, we identify four research directions in the field of DNS encryption yet to be investigated.
\end{itemize}

\subsection{Roadmap}
The organization of this paper is depicted in Fig.~\ref{fig:keyLogistic}.
We discuss the current development of DNS encryption in \S\ref{sec:SurveyDevelopment}, malware communications leveraging DNS encryption in \S\ref{sec:SurveyMalware}, and inference techniques for encrypted DNS traffic in \S\ref{sec:SurveyDetection}.
We also list the references cited in each of the above three sections in Table~\ref{tab:paperCategory}.
Four future research directions are presented in \S\ref{sec:discussion}.
Related surveys on DNS security are discussed in \S\ref{sec:relatedWork}. This survey is concluded in \S\ref{sec:conclusion}.

\section{DNS Encryption: Methodologies, Adoption \& Performance, Benifits, and Issues}\label{sec:SurveyDevelopment}

Encrypting DNS queries and responses between clients and resolvers holds the promise of protecting user privacy against eavesdroppers and man-in-the-middle (MITM) attackers \cite{PWu2019}. In this section, we review the development of DNS encryption by highlighting current standard techniques (\S\ref{sec:method}), their adoption status and performance impacts (\S\ref{sec:adoption}), benefits provided by DNS encryption (\S\ref{sec:benefit}), and its open issues and security vulnerabilities (\S\ref{sec:secIssues}).

\begin{figure*}[!t]
	\begin{center}
		\includegraphics[width=1\textwidth]{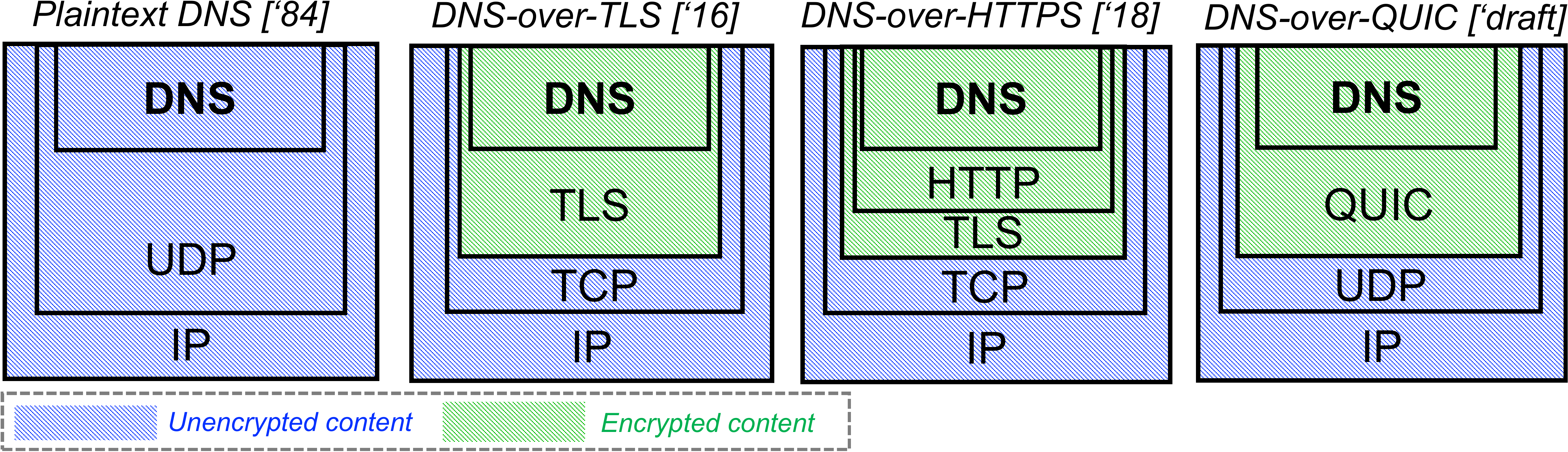}
		\caption{The structure of packet layer (starting from IP layer) of plaintext DNS, encrypted DNS over TLS, encrypted DNS over HTTPS, and encrypted DNS over QUIC.}
		\label{fig:DNSpacket}
		\vspace{-3mm}
	\end{center}
\end{figure*}

\subsection{Standard Methods of DNS Encryption}\label{sec:method}
In general, DNS encryption is achieved by encapsulating the content of queries and responses (between clients and resolvers) in an upper layer protocol that encrypts the  packet content using available cryptographic techniques.
To date, there are three standardized DNS encryption methods, each utilizing a specific upper-layer protocol, including DNS-over-TLS (DoT), DNS-over-HTTPS (DoH), and DNS-over-QUIC (DoQ), ordered by their time of proposal.
As illustrated in Fig.~\ref{fig:DNSpacket}, unlike the plaintext-based DNS that is directly embedded under the UDP transport layer, the three DNS encryption methods encapsulate DNS in their encryption-enabled layer, \ie TLS, TLS with HTTP (\ie HTTPS), and QUIC, respectively.

\subsubsection{DNS over TLS (DoT)}
This is the first standard protocol proposed for DNS encryption  that has its RFC \cite{rfcDNSTLS2016} published in 2016. It uses a Transport Layer Security (TLS) layer under the TCP transport layer to encrypt DNS queries and responses. Before being applied to the DNS applications, TLS has proven its efficacy in encrypting popular network applications such as email (SMTP), hypertext (HTTP), and voice-over-IP (VoIP).

During lookups, the client first initiates a TCP connection to a designated DoT port {\myverb{TCP/853}} on its intended DoT-enabled resolver. Next, a TLS connection will be established via a typical TLS handshake process to exchange their cryptographic keys \cite{rfcTLS2008}. Upon successful establishment of the TLS session, the client is able to perform TLS-encrypted DNS lookups through the DoT port {\myverb{TCP/853}} on the resolver side.
Depending on the configurations of clients and servers, the TLS connections may remain open for further DNS lookups, reducing latency (\ie preventing additional TCP/TLS handshakes for subsequent requests). 

Although DoT is a viable approach for DNS encryption, it faces several practical challenges that may limit its usage.
The first concern is that, as reported in \cite{RKartch2017}, the port {\myverb{TCP/853}} dedicated to DoT services is currently not well recognized by the security community thus is likely to be blocked by firewall appliances.
While the content of queries and responses is encrypted, eavesdroppers can obtain encrypted DNS packets relatively easier (compared to DoH and DoT where a mix of traffic is exchanged) by filtering TCP/853 and performing statistical analysis (\eg on packet sizes) to infer their embedded contents (will be discussed in \S\ref{sec:SurveyDetection}).
Another shortcoming of DoT is that it requires application developers and hardware manufacturers to support the protocol -- if not supported, users may still go unprotected.
Note that DoT is less resilient to packet losses \cite{DStenberyCCR2014} (suffers from  head-of-line blocking) compared to its counterparts, DoH and DoQ (discussed next). In addition to this technical shortcoming, the timing of when DoT was proposed seems relatively early (compared to DoH, for example) -- it was introduced when encryption was not very commonplace.
The above mentioned facts have negatively impacted the public adoption of DoT protocol.

\subsubsection{DNS over HTTPS (DoH)}

DoH \cite{rfcDNSHTTPS2018} has been standardized in 2018 to address the practical difficulties of DoT discussed above. Anyone using a supported web browser automatically benefits from encrypted DNS.
It utilizes the widely-used HTTPS protocol (\ie TLS with HTTP as depicted in Fig.~\ref{fig:DNSpacket}) to encapsulate DNS contents, delivered via the service port {\myverb{TCP/443}} so that existing security measures would not hinder its transmission.
DoH is also more acceptable on the client-side as HTTPS is the default (or the only enabled) hypertext protocol used by major browsers. From a privacy viewpoint, DoH is more preferred than DoT since DNS traffic is mixed with other applications (\eg Web) exchanged over HTTPS.
In addition, any future performance and security enhancements for HTTPS would also benefit DoH.
Given the above practical benefits, DoH has become the most popular DNS encryption method used by the current Internet industry \cite{DFisher2020}.

\subsubsection{DNS over QUIC (DoQ)}

The transport-layer protocol of both DoT and DoH is TCP.
Considering TCP and TLS handshakes required for connection establishment plus the TCP acknowledgement mechanism, a DNS request of DoT and/or DoH can take longer (compared to plaintext UDP-based) to get responded. Such overhead is often non-negligible
Therefore, DoQ \cite{rfcDNSDoQ2021} was first proposed in 2017 by Google as an Internet draft (not officially finalized as an RFC yet) to further improve the performance of encrypted DNS.
It leverages QUIC \cite{rfcQUIC2021} protocol (built on the top of UDP transport layer) that enables faster and lighter encrypted communications by zero-RTT handshakes \cite{AGhedini2019} and multiplexing data streams. 
As reported by G. Carlucci \textit{et al.} \cite{GCarlucciSAC2015}, QUIC empirically outperforms its competitor HTTPS in response quality metrics like page load times.

Despite its performance merits, similar to DoT, DoQ uses dedicated service ports ({\myverb{UDP/784}} and {\myverb{UDP/8853}}) that are not well-recognized by security systems at present time, therefore, its traffic might get blocked (as the default option) by firewalls during transmission.

\subsection{Adoption and Performance of DNS Encryption}\label{sec:adoption}
Having understood the current DNS encryption techniques and protocols, we now discuss their state of adoption across the Internet as well as their performance reported by the data networking community.

\subsubsection{Current State of Public Adoption}
As described in \S\ref{sec:method}, DNS encryption is applied for the queries and responses between clients and resolvers, and hence, the public adoption highly depends on the technical supports from two key stakeholders, \ie public resolvers and user applications (\eg browsers and operating systems).

\textbf{Public Resolvers:} A few public resolvers, mostly operated by major cloud providers such as Google and Cloudflare, start to support DNS encryption since 2019 \cite{MVale2019}, followed by many other providers such as AdGuard, Alibaba, Cisco, Comcast, and Quad9. 
According to two technical reports \cite{DNSEncryptedResolver, DoHResolver}, there are at least seven DoT-enabled and 62 DoH-enabled public resolvers on the Internet. 
AdGuard announced \cite{VBagirov2020} the support of DoQ on their public resolvers since December 2020, though this protocol is still in its experimental stage.

The public adoption of DNS encryption holds a growing promise, evidenced by its increasing Internet-wide traffic volumes \cite{SGarciaARXIV2021}. However, it is dominated by a small number of major service providers, raising concerns of data monopolization \cite{MCWasastjernaECJ2018}.
Also, there are oppositions, like the UK Internet Services Providers’ Association (ISPA) \cite{TANideck2019}, who argue that DNS encryption can bypass existing filtering obligations, thus undermine the Internet safety standards.

\begin{table*}[!t]
	\caption{DNS encryption options provided by user applications.}
	\centering
	\normalsize
	\renewcommand{\arraystretch}{1.2}
	\begin{tabular}{|l|p{10cm}|}
		\hline
		\rowcolor[rgb]{ .906,  .902,  .902} 	\textbf{Option}              & \textbf{Description} \\ \hline
		Off           & Plaintext without attempting to apply encryption for DNS communications. \\ \hline
		Opportunistic/Automatic \cite{TBucutiISI2015}                   &   Apply encryption if possible by both client and server; otherwise, plaintext DNS (if no encryption protocol is available/possible).\\ \hline
		Strict & Apply specified encrypted DNS protocols. \\ \hline
	\end{tabular}
	\label{tab:DNSMode}
\end{table*}

\textbf{User Applications:}
DNS encryption is increasingly supported by a variety of end-user platforms and applications including operating systems for mobile and desktop devices, Internet broswers, and even IoT devices \cite{CHesselmanIC2020}. 
As reported by \cite{PWu2019}, major operating systems including Android, iOS, Linux, macOS, and Windows have included DoH features in their latest versions. 
Similarly, popular Internet browsers such as Firefox, Chrome, and Opera also offer DNS encryption options to their users -- ``Off'', ``Opportunistic'' (or ``Automatic''), and ``Strict''.
By the ``Off'' mode, DNS lookups will be performed via plaintext. For the ``Opportunistic'' mode, a client and its resolver try to agree on a DNS encryption protocol that is available on both sides. However, plaintext data exchange is used if no agreement can be made. For the ``Strict'' mode, DNS lookups can only be performed via a certain encryption protocol specified by the user. Their short descriptions are also summarized in Table~\ref{tab:DNSMode}.

However, according to an online survey studying the attitude of clients towards DNS encryption \cite{ANisenoffDNSPW2021}, users are quite reluctant and reactive to accept DNS encryption.
Majority (67\%)  of the surveyed population selected the Opportunistic option followed by Strict (20\%) and Off (13\%) options.
This indicates that the adoption of DNS encryption would be highly contingent on how easily application developers enable DNS encryption features on their application, without additional manual configurations from users.
One such early example is that Firefox started to bring DoH as its default DNS protocol for their U.S. based users since 2020 \cite{SDeckelmann2020}.

\subsubsection{Performance Analysis}\label{sec:performanceAnalysis}
Compared with UDP-based DNS in plaintext, encrypted DNS requires additional communication steps for exchanging cryptographic keys (\eg TLS handshake) and TCP communications (\eg TCP handshake and payload acknowledgment). Therefore, introducing encryption to DNS inevitably brings more computational and communication overheads that might not be favorable to sensitive applications (\eg website fetching) \cite{ZYanFGCS2020,HShulmanWPES2014}.
A number of research articles have attempted to quantify the performance implications of DNS encryption by empirically measuring various metrics such as page load time, DNS resolution time, or query response time across a range of access network types (\eg cellular, campus, or home networks), and resolvers. Their setup details are summarized in Table~\ref{tab:performanceSetup} and will be explained next.

\begin{table*}[!t]
	\caption{A summary of measurement setup in prior works on performance analysis of DNS encryption.}
	\centering
	\normalsize
	\renewcommand{\arraystretch}{1.2}
	\begin{tabular}{|l|p{2cm}|p{3cm}|p{3.2cm}|p{4cm}|}
		\hline
		\rowcolor[rgb]{ .906,  .902,  .902} 	\textbf{Work}           & \textbf{Protocols}   & \textbf{Key metrics} & \textbf{Network conditions} & \textbf{Target resolvers} \\ \hline
		\cite{AHounselANRW2019} & DNS, DoH, DoT & page loading time& 3G, 4G, campus networks & Cloudflare \\ \hline
		\cite{TBottgerIMC2019} & DNS, DoH, DoT & query resolution time, page loading time& not specified&  local resolver, Google, Cloudflare \\ \hline
		\cite{KBorgolteEJ2019} & DNS, DoH & page loading time& 4G, campus networks & Google, Cloudflare, Quad9 \\ \hline
		\cite{AHounselWWW2020} & DNS, DoH, DoT & query resolution time, page loading time& five different ISP networks & Google, Cloudflare, Quad9 \\ \hline
		\cite{ESMbeweAFRICOMM2021} & DNS, DoH, DoT & query resolution time, page loading time& mobile, community, and educational networks & local and five public resolvers\\ \hline	
		\cite{TNDoanPAM2021} & DNS, DoT & query resolution time, failure rate & home networks (mainly in NA \& EU) & one local and 15 public resolvers  \\ \hline	
		\cite{AHounselPAM2021} & DNS, DoH, DoT & query resolution time, latency & home networks in MBA program  & three anonymized public resolvers \\ \hline	
		\cite{RChhabraIMC2021} & DNS, DoH & query resolution time, number of requests per connection & clients from 2190 different autonomous systems & Cloudflare, Google, Quad9, NextDNS\\ \hline	
	\end{tabular}
	\label{tab:performanceSetup}
\end{table*}

A. Hounsel \textit{et al.} \cite{AHounselANRW2019} studied the impact of DNS options (plaintext DNS, DoH, and DoT) on the load time of web pages in various networks, including cellular 3G, lossy cellular 4G, or campus wired networks.
Their results show that page load times of the three protocols under the (near ideal) university network are almost identical, as their statistical differences approach 0 seconds. Under the 3G network, plaintext DNS performs the best, followed by DoT and DoH. However, DoT gives the best performance in page load time, followed by DoH and plaintext DNS in the lossy 4G networks. The authors believe that possibly it is because TCP (employed by DoT and DoH) has shorter timeout thresholds than UDP (employed by plaintext DNS).

T. Bottger \textit{et al.} \cite{TBottgerIMC2019} quantified the overhead of DNS encryption in resolution time and page load time introduced by the TLS layer (in DoT and DoH) and HTTP layer of versions 1.1 and 2.0 (in DoH). They concluded that both DoT and DoH via HTTP1.0 have significant performance degradation (\eg additional delays from more than 0.1 to 1 second) due to the head-of-line blocking problem, while DoH using HTTP2.0 is more promising as it results in similar delays compared with plaintext DNS.

K. Borgolte \textit{et al.} \cite{KBorgolteEJ2019} investigated the page load time of DoH and plaintext DNS under 4G and campus network scenarios using three different open resolvers by Google, Cloudflare, and Quad9. The authors revealed that only the DoH service of Cloudflare under the university network gives the most reasonable load time. At the same time, other combinations incur significant delays of up to several seconds (\eg about 5 seconds for DoH using Quad9 resolver under the 4G network). 
The authors also identified that lossy network conditions and sub-optimal provider selections might widen the performance gaps between DNS and DoH.

A. Hounsel \textit{et al.} \cite{AHounselWWW2020} highlighted that although plaintext DNS has better performance in query resolution time (\ie about 300ms and 450ms faster than those of DoT and DoH, respectively), page load times of DoT and DoH are better than that of plaintext DNS (\ie 101ms and 33ms faster, respectively) in the lossy 4G network. Possibly, this is because it takes a longer time for UDP (plaintext DNS) to detect a lost query, while TCP can quickly react and transmit the lost packet over an existing connection. Given its connection-less property, UDP achieves a shorter response time than TCP in ideal situations (lossless network).
In addition, TCP and HTTPS connections initiated by DoT and DoH lookups could be reused by the following web browsing activities to reduce the time overhead introduced by TCP and TLS handshakes.
Also, suppose the network becomes heavily lossy (\eg a 3G network in a remote location). In that case, UDP-based DNS outperforms TCP-based versions (DoH and DoT) since the time for connection establishment would dominate, and UDP would have time to react to losses.

From the above studies, it is quite clear that DNS encryption protocols perform not so well when network conditions are non-ideal. A similar insight was drawn from measurement studies on edge networks \cite{RChhabraIMC2021,ESMbeweAFRICOMM2021,TNDoanPAM2021,AHounselPAM2021} (\eg home) that are far from the essential Internet infrastructures.
By measuring DoH resolution times from 22K unique hosts worldwide, 
R. Chhabral \textit{et al.} \cite{RChhabraIMC2021} found that hosts in high-income countries/regions with better Internet infrastructure are less likely to have performance degradation.
At the same time, clients from less-developed areas experience a significant slowdown by switching from plaintext DNS to DoH.

E. S. Mbewe \textit{et al.} measured the performance of DoH, DoT, and plaintext DNS of hosts in Africa \cite{ESMbeweAFRICOMM2021} and observed high latency and circuitous DNS resolution paths for encrypted DNS packets. For example, round-trip times from South Africa are around 150ms, while the values for Madagascar and Uganda are more than 1000ms.

T. N. Doan \textit{et al.} \cite{TNDoanPAM2021} measured DoT lookups from 3.2K RIPE Atlas probes deployed in home networks. The author first pointed out that only 0.4\% of the studied households have their local resolvers supporting DoT. Through extensive measurements, they found higher failure rates (up to 32\%) and response times (\eg about 130ms to 230ms) in DoT than in plaintext DNS, where they observe much lower failure rates (less than 3\%) and faster response (less than 100ms).

A. Hounsel \textit{et al.} \cite{AHounselPAM2021} performed measurements on more than 2500 home networks participating in a Measuring Broadband America program funded by the Federal Communications Commission (FCC). They highlighted that DNS clients could periodically conduct active probing in order to select their optimal user settings in terms of the choice of protocol (DoT or DoH) and resolvers that give a satisfying performance for each home network.

\subsection{Benefits of DNS Encryption}\label{sec:benefit}
Regardless of some performance issues, the industry has advocated encrypted DNS due to its unprecedented benefits such as protecting user privacy and preventing certain attacks that exploit the connectionless property of UDP.

\subsubsection{Protecting User Privacy}
The risk of end-user privacy leakage is present in various nodes across the path of DNS lookups, shown in Fig.~\ref{fig:DNSLookup}. 
As highlighted by H. Shulman \textit{et al.} \cite{HShulmanWPES2014} and Z. Yan \textit{et al.} \cite{ZYanFGCS2020}, eavesdroppers may monitor:
\begin{itemize}
	\item DNS queries and responses through the links between clients and recursive resolvers.
	\item DNS logs from recursive resolvers.
	\item DNS queries and responses through the links between recursive resolvers and authoritative name servers.
	\item DNS logs from authoritative name servers.
\end{itemize}

Among these four risks, the first one is of the highest importance since both the client identity and the domain name are exposed to malicious actors who perform passive traffic sniffing. 
In addition, retrieving logs from recursive resolvers (the second risk listed above) would result in the same outcomes for adversaries; however, it requires them to compromise the servers.
On the other hand, plaintext DNS lookups between resolvers and name servers can only reveal an aggregate profile of (\eg millions of) users served by the public recursive resolver without exposing the identity of individual users.
Also, the logs from authoritative name servers do not give away individual user information.
DNS encryption protocols, therefore, are primarily designed to protect queries and responses from third-parties, eavesdropping between clients and recursive resolvers \cite{KBumanglagICICT2020}, as decrypting ciphertext would be largely impractical without knowledge of the secret key.

\subsubsection{Preventing Network Attacks on UDP-based DNS}
In addition to preserving user privacy, applying encryption techniques to DNS traffic requires the establishment of a secure connection (\eg TLS, HTTPS, or QUIC) between the client and intended resolver. It changes the connectionless nature of legacy (plaintext) DNS exchanged via port {\myverb{UDP/53}} that may be exploited by malicious actors for various types of network attacks (discussed next). Therefore, applying encryption prevents certain network attacks in the DNS ecosystem.
For example, \textbf{\textit{DNS amplification attack}} is a popular form of distributed denial of service (DDoS) that relies on the use of publicly accessible open DNS servers to overwhelm a victim system with DNS response traffic \cite{MAnagnostopoulosCS2013}. Attackers often craft small-sized DNS queries with source IP addresses spoofed to be a victim’s address and send them to resolvers. When resolvers send the DNS record response, it is sent instead to the victim (never requested anything). Attackers will typically submit a request for as much zone information as possible to maximize the amplification effect. Because the size of the response is larger than the request, the attacker is able to increase the amount of traffic directed at the victim.
Such attacks are hard to be prevented by the resolvers if DNS queries arrive over connectionless protocol -- resolvers are unable to determine whether a query is with spoofed source IP addresses or not.
DNS encryption protocols, instead, prevent this problem by requiring a connection to establish via handshaking processes \cite{LZhuSP2015}. Therefore, no large-sized DNS response could be sent to a (spoofed) victim IP address.

In addition, L. Zhu \textit{et al.} \cite{LZhuSP2015} demonstrated that \textbf{direct DDoS attack on DNS servers} is weakened by DNS encryption. This is because attackers need to establish TCP and TLS connections for each attempt of attack (\ie malicious DNS lookup), leading to higher computing resources required to overwhelm a DNS server.
Lastly, encrypting DNS communications between clients and resolvers also reduces the chance of \textbf{man-in-the-middle (MITM) DNS spoofing attacks} \cite{NNayakICCSIT2010} that aim to mislead (\ie redirect) clients towards malicious destination IP addresses by hijacking and manipulating DNS responses.
As a use-case, P. Jeitner \textit{et al.} \cite{PJeitnerDSNS2020} developed a secured addressing mechanism using DoH for network time protocol (NTP) systems to prevent the off-path attacks \cite{JPhilippDSN2020} which redirect clients to malicious timing servers via manipulated DNS responses.

\begin{table*}[!t]
	\caption{Practical issues and security vulnerabilities of encrypted DNS highlighted by existing literature along with impacted  protocols as discussed in \S\ref{sec:secIssues}.}
	\centering
	\normalsize
	\renewcommand{\arraystretch}{1.2}
	\begin{tabular}{|p{6.3cm}|p{4.9cm}|p{2.2cm}|}
		\hline
		\rowcolor[rgb]{ .906,  .902,  .902} 	\textbf{Practical issues and security vulnerabilities}              & \textbf{References} & \textbf{Protocols}\\ \hline
		Privacy leakage by compromised resolvers (\S\ref{sec:secIssues1})          & \cite{LJinWWW2021,SRiveraARES2020,YNakatsukaACSAC2019,YNakatsukaDT2021} & DoT, DoH, DoQ\\ \hline
		Invalid SSL certificates     (\S\ref{sec:secIssues2})             & \cite{CLuIMC2019,jahromi2021comparative} & DoT, DoH \\ \hline
		Fallback attacks (\S\ref{sec:secIssues3})    & \cite{DShinACSAC2011,QHuangFOCI2020} & DoT, DoH \\ \hline
		Imperfect padding strategies (\S\ref{sec:secIssues4})    & \cite{RHouserCoNEXT2019,rfcDNSPadding2018,KHynekIEMCON2020,RHouserCoNEXT2019,JBushartFOCI2020} & DoT, DoH\\ \hline
		Problems of monopolized DNS resolution (\S\ref{sec:secIssues5})    & \cite{KRadinsky2015,PSchmittANRW2019,rfcODNS2021,AHounselANRW2021} &DoH \\ \hline
		Existing barriers for public adoption (\S\ref{sec:secIssues6})    & \cite{MLyuPAM2019,JAhmedTNSM2020,MLyuTNSM2021,DoHThreatLandscape,PBischoff2021,ZYanFGCS2020, HHabibiGharakheiliCAN2017,AJMartin2019,CCimpanu2019,hynek2020doh,MSinghDI2019,KAlieyanNCA2017} & DoT, DoH, DoQ \\ \hline
	\end{tabular}
	\label{tab:SecurityIssue}
\end{table*}

\subsection{Practical Issues and Security Vulnerability}\label{sec:secIssues}
As highlighted by recent research, although DNS encryption has brought significant enhancements in security and privacy, there are still unsolved issues and security vulnerabilities of encrypted DNS. Some of those include privacy leakage by compromised resolvers, invalid SSL certificates, fallback attacks, imperfect padding strategies, problems of monopolized DNS resolution, and existing barriers for public adoption which are discussed as follows. 
Table~\ref{tab:SecurityIssue} summarizes these issues and vulnerabilities along with their respective prior research and impacted protocols. Note some protocols may truly be subject to a certain category but not listed since they have not been highlighted (reported) by the existing literature. This particularly applies to DoQ, which is relatively a recent encryption protocol compared to other counterparts.

\subsubsection{Privacy leakage by Compromised Resolvers}\label{sec:secIssues1}
As discussed in \S\ref{sec:benefit}, protocols like DoT, DoH, and DoQ focus on encrypting DNS communications between clients and a recursive resolver, primarily aiming to protect user privacy. Unlike authoritative name servers, recursive resolvers would have the ability to construct a comprehensive profile of users who primarily aimed to keep it private by way of DNS encryption. One may argue that resolver compromise exists with plaintext DNS and continues to exist with encrypted DNS. We note protecting user privacy was not an objective in plaintext DNS, but it is for encrypted DNS protocols. That’s why the security of resolvers is of particular interest in the context of encrypted DNS since malicious actors may obtain data of user activities by compromising a recursive resolver.
According to L. Jin \textit{et al.} \cite{LJinWWW2021}, payload manipulation by compromised resolvers is not uncommon even for encrypted DNS.
The authors performed more than seven million lookups towards thousands of DoT/DoH-enabled DNS resolvers. They observed that more than 1.5\% of responses are manipulated according to their ground-truth records of domain names and the respective IP addresses.

To tackle this potential threat, S. Rivera \textit{et al.} \cite{SRiveraARES2020} developed a privacy-preserving mechanism that leverages extended Berkeley Packet Filter (eBPF) to assist users in distributing their queries towards a set of resolvers randomly. That way, a compromised resolver would not expose the entire history of query records from a client.
Y. Nakatsuka \textit{et al.} \cite{YNakatsukaACSAC2019,YNakatsukaDT2021} developed PDoT, an architecture that hosts recursive resolvers in a Trusted Execution Environment (TEE), so that they can be verified, authenticated, and trusted by end-users before performing DNS lookups.

\subsubsection{Invalid SSL Certificates}\label{sec:secIssues2}
The SSL certificates of DoT and DoH resolvers are provided to users as proof of their identities prior to the start of encrypted sessions (the process is known as SSL handshake).
As reported by C. Liu \textit{et al.} \cite{CLuIMC2019}, a non-negligible fraction (25\%) of 150 DoT servers they studied have invalid SSL certificates that may impose privacy risks to clients as they cannot be verified as trusted entities.
Similarly, A. S. Jahromi \textit{et al.} \cite{jahromi2021comparative} analyzed about $10K$ SSL certificates collected from public DoT servers and found out that only 65\% of them are checked as valid.
In contrast, the popular reasons for invalidity include non-existent issuers, self-assigned certificates, expired certificates, and expired windows.
Note that some of the resolvers with invalid SSL certificates may not be intended for public usage – they may be abandoned servers left from short-term projects on a university network. In summary, given observations from real-world measurements, it is important for service operators to maintain valid certificates when offering encrypted DNS services to the public.

\subsubsection{Fallback Attacks}\label{sec:secIssues3}
According to \S\ref{sec:adoption}, clients often use the ``opportunistic'' (or ``automatic'') mode as their default option to establish encrypted connections with resolvers via the available DNS encryption protocol. Plaintext-based DNS will be used if both sides agree on no encryption protocol.
Unfortunately, malicious actors may utilize this fact to steal user privacy by forcing the clients to perform their DNS lookups in plaintext -- it is known as a fallback (or rollback/downgrade) attack \cite{DShinACSAC2011}.
Fallback attacks exploit a feature offered by servers operating updated/secured versions of protocols (\eg DoH and DoT) that may allow a client to communicate via older version counterparts (\eg plaintext DNS) for backward compatibility. During attacks, a malicious actor often creates fake negotiations (\eg through man-in-the-middle) between the client and server so that a less-secure protocol is used in the following communications.

As for the viability of fallback attacks, Q. Huang \textit{et al.} \cite{QHuangFOCI2020} reported that four techniques could achieve such fallback, including DNS traffic interception, DNS cache poisoning, TCP traffic interception, or TCP reset injection. The authors have experimented with the four methods on six major browsers such as Chrome and Firefox. Although the response patterns exhibit variations, all tested browsers are found vulnerable to at least one of the four fallback techniques. Therefore, the authors suggested application developers revisit their encryption policies (\eg notifying users if the plaintext is chosen) instead of simply leaving ``opportunistic'' as the default option.

\subsubsection{Imperfect Padding Strategies}\label{sec:secIssues4}
DNS encryption seems to be promising in protecting privacy by way of ciphertext. However, a recent research \cite{RHouserCoNEXT2019} found that eavesdroppers can still infer the visited websites of clients with more than 99.5\% accuracy through statistical features of DoT traffic such as the temporal patterns of packet sizes.

To prevent data leak from encrypted DNS messages via size-based traffic analysis, the Internet Society proposed a padding strategy in 2018 \cite{rfcDNSPadding2018} that fills the DNS queries and responses to a certain size as configured by the users. 
However, according to K. Hynek \textit{et al.} \cite{KHynekIEMCON2020}, DNS message padding was not well supported (at least until 2020) by the majority of browsers such as Firefox, thus, its public adoption is inevitably limited.

Regardless of its public adoption, padding encrypted DNS packets is not a perfect solution against data leakage.
R. Houser \textit{et al.} \cite{RHouserCoNEXT2019} managed to achieve more than 80\% accuracy in inferring visited websites by analyzing the temporal patterns of padded DoT messages.
J. Bushart \textit{et al.} demonstrated in \cite{JBushartFOCI2020} how their method classifies (more than 85\%) client device types and predicts (more than 65\%) visited domains by combining the size and inter-arrival timing information of padded DoT and DoH from end-users.

\subsubsection{Problems of Monopolized DNS Resolution}\label{sec:secIssues5}
In the current ecosystem of encrypted DNS, a few major providers such as Google and Cloudflare dominate the market of public resolvers on the Internet. Instead, in plaintext DNS systems, recursive resolvers are well distributed across Internet service providers (ISP) and large public organizations.
As discussed next, the related articles pointed out that such a monopolization (also known as centralization) can be detrimental to the reliability and usability of encrypted DNS resolution.

First, having a limited number of major providers dominating the ecosystem can lead to data monopolization \cite{KRadinsky2015}, which can threaten users' privacy as they would be able to profile user online preference and behavior by their queried domain names. Apart from having more players in this ecosystem, researchers have developed novel methods like Oblivious DNS (ODNS) \cite{PSchmittANRW2019} to alleviate this problem.
P. Schmitt \textit{et al.} \cite{PSchmittANRW2019} introduced a client proxy to operate between clients and public resolvers -- client proxies receive queries from end-hosts and send them to respective recursive resolvers without revealing their identities. Therefore, with ODNS, monopolized service providers would not have the IP address of their clients. In 2021, ODNS over HTTPS (ODoH) has been standardized as an RFC \cite{rfcODNS2021} to prevent the potential misuse of user data by resolver providers. We believe that it is an effective method yet requires public awareness before it is widely adopted.

Second, the monopolized DNS resolution via encrypted protocols provides a sub-optimal quality of service (\eg resolution time) for user devices that require flexibility in setting their strategy for dynamic selection (depending on geolocation and network conditions) of recursive resolvers.
To this end, A. Hounsel \textit{et al.} \cite{AHounselANRW2021} developed a refactored DNS resolver architecture that supports de-monopolized name resolution, enabling users to specify their lookup preference such as distributing queries across a set of DoH resolvers with certain latency requirements -- provides an optimal and flexible selection of resolvers for edge devices. Their prototype evaluation demonstrated an improved performance while preserving user privacy.

\subsubsection{Existing Barriers for Public Adoption}\label{sec:secIssues6}
The encryption of DNS queries and responses significantly impacts the usability of legacy security measures that inspect plaintext DNS payload for inference and classification purposes. According to the current literature, we now enumerate some security functionalities that have been affected by DNS encryption.

\begin{itemize}
	\item \textbf{Enterprise security via DNS monitoring:} 
	Legacy middleboxes for network security (\eg border firewalls and network intrusion detection systems) can be configured to inspect DNS messages, gaining visibility into the role of connected assets \cite{MLyuPAM2019} and/or detecting anomalous/malicious patterns \cite{JAhmedTNSM2020,MLyuTNSM2021}, indicative of DNS-based volumetric attacks, accessing illegal contents, or performing data exfiltration. 
	However, payload inspection becomes relatively ineffective with the increasing adoption of encrypted DNS. Hence, new challenges  \cite{DoHThreatLandscape} such as lack of complete visibility into DNS traffic, losing control over DNS data, potential leakage of information, and inability to block illegal (inappropriate) contents emerge for enterprise network operators.
	
	\item \textbf{Internet censorship, parental control, and advertisement blocking:} 
	Internet censorship has been enforced at various levels by the government of many countries such as the U.S., China, and the U.K. \cite{PBischoff2021} to restrict access to certain online services such as pornography, violence, and scandals.
	Also, monitoring DNS traffic has been used for home networks for parental control purposes \cite{ZYanFGCS2020, HHabibiGharakheiliCAN2017}.
	As criticized by relevant stakeholders such as the Internet Watch Foundation \cite{AJMartin2019}, with encrypted DNS such as DoH, users can easily bypass censorship imposed at home, organizational, ISP, or national level \cite{CCimpanu2019}.
	In addition, the efficacy of ad-blocking mechanisms, which rely on blocklists of DNS names, is impacted  \cite{hynek2020doh}.
	
	\item \textbf{DNS-based criminal investigations:} 
	Internet service providers \cite{ZYanFGCS2020} are often required by laws to record online activities (queried domains) of their subscribers for a certain period that may be needed for criminal investigations. 
	Therefore, serious concerns are raised by agencies like Government Communications Headquarters (GCHQ) that encrypted DNS (as it bypasses current surveillance systems) would cause significant challenges for lawful investigation \cite{CCimpanu2019}.

	\item \textbf{Captive portal via DNS hijacking:}
	Commercial venues (\eg shopping malls and airports) that offer public wireless Internet access to customers often use captive portals to manage their network access through their network gateways. Via DNS hijacking, unauthorized users will be redirected to the login page configured by operators regardless of their actual destinations specified in DNS lookups. The adoption of DNS encryption may render the captive portals ineffective \cite{ZYanFGCS2020} as the network gateways are no longer able to modify DNS responses sent to the end users.
	
	\item \textbf{Detection of malware distribution:} DNS plays a critical role in the activities of malware. Therefore, security experts have developed methods to detect malware distribution by analyzing domain names in plaintext DNS messages \cite{MSinghDI2019,KAlieyanNCA2017}. Unfortunately, those methods are incapable of processing encrypted DNS packets, which do not give visibility into the DNS message content (\eg query names). This topic will be discussed in detail in \S\ref{sec:SurveyMalware}.
\end{itemize}

\begin{figure*}[!t]
	\begin{center}
		\includegraphics[width=1\textwidth]{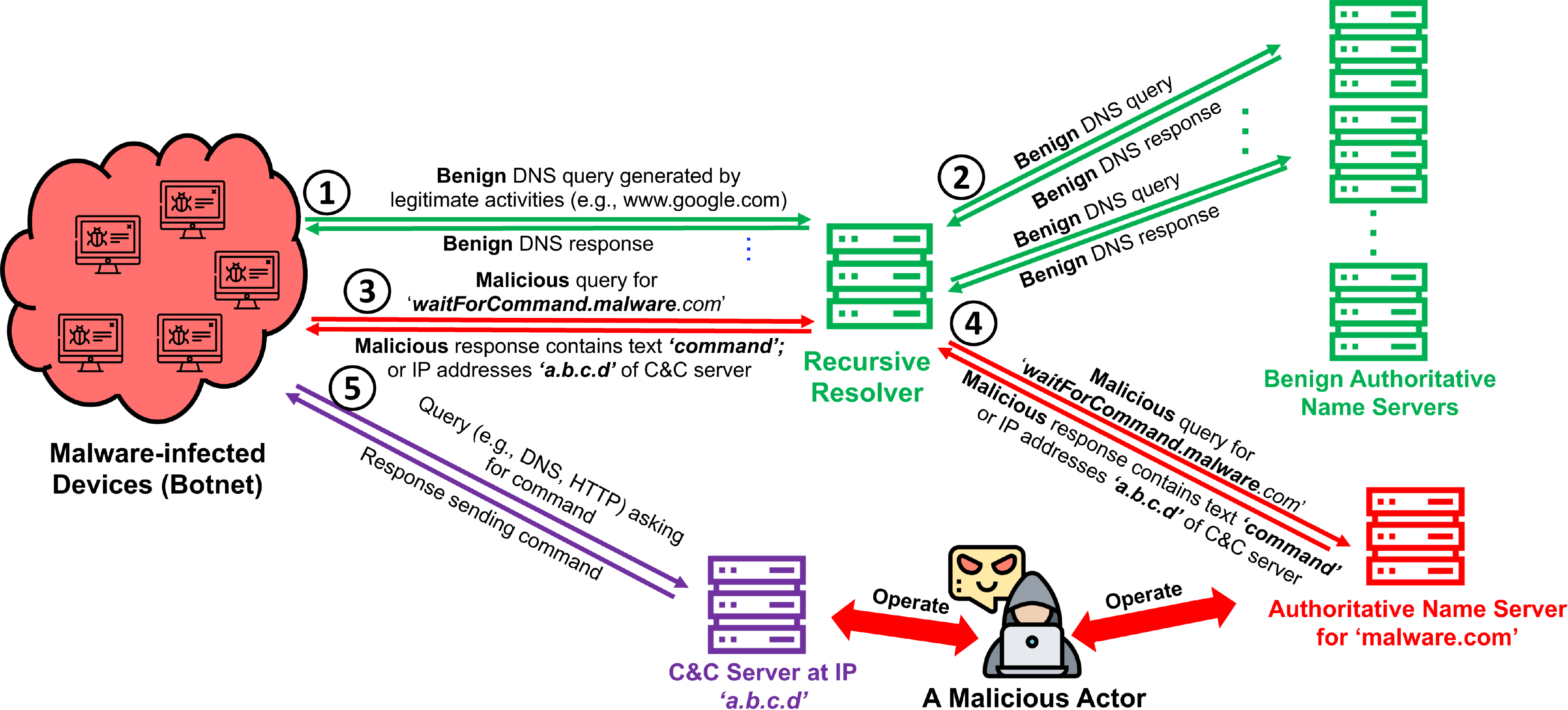}
		\caption{A visual example of command-and-control (C\&C) communications via DNS from malware-infected devices.}
		\label{fig:CC}
	\end{center}
\end{figure*}

\section{Malware Communications Empowered by DNS Encryption}\label{sec:SurveyMalware}

Various malware families frequently use DNS for command-and-control (C\&C) communications and data exfiltration \cite{CPatsakisCS2020,Godlua2019,PsiXBot2019}. While the security community has comprehensively understood such misuse via plaintext DNS, the threat landscape introduced by encrypted DNS has not been well studied yet. This section reviews some of the known ways of malware misusing encrypted DNS by malware and highlights emerging challenges for security teams for detecting those malicious activities. We focus on C\&C communications (\S\ref{sec:CC}) and data exfiltration (\S\ref{sec:Exfiltration}).
\subsection{Command-and-Control (C\&C) Communications}\label{sec:CC}

Malicious actors on the Internet spread their malware to infect connected devices by various methods such as phishing emails, scripts embedded in web pages, and manipulating vulnerable firmware. Those malware-infected devices form botnets that establish connections with the external malicious actors to conduct cyber crimes as instructed automatically. The process of establishing connections between botnet and masters followed by exchanging information and instructions between these two parties is known as command-and-control (C\&C) communications.

\subsubsection{C\&C Communications via Plaintext DNS}\label{sec:CCPlaintext}
Security appliances and middleboxes (in most enterprise networks) often permit DNS traffic \cite{MLyuPAM2019}. Attackers, therefore, take advantage of getting free rides and make C\&C communications via DNS instead of other proprietary protocols to bypass possible security measures that may block malicious traffic \cite{CommandAndControl}.

\textbf{What is C\&C over DNS?:}
Fig.~\ref{fig:CC} depicts a typical process of DNS-based C\&C communications. In parallel to benign DNS lookups through recursive resolvers (step \ciao{1} and \ciao{2}) for legitimate activities, malware-infected devices send DNS queries with intentions embedded in the questioned domain names (\eg {\myverb{waitForCommand.malware.com}}) towards a malicious authoritative name server (\eg for {\myverb{malware.com}}) operated by an adversary (step \ciao{3} and \ciao{4}). Note that malicious queries are transmitted to typical DNS recursive resolvers, much like the way for benign DNS lookups.
The authoritative name server for {\myverb{malware.com}} may directly send instructions to its botnet via DNS responses with text format (\ie TXT type). Also, as reported by \cite{KXuTDSC2013}, it is common for other types of C\&C communications (\eg HTTP-based) to locate their masters via DNS resolution, \ie sending the IP address of a dedicated C\&C server ({\myverb{a.b.c.d}} in step \ciao{5}) for the malware-infected device to connect.

\textbf{What are the countermeasures against C\&C over plaintext DNS?:}
Malware-infected devices that perform regular C\&C communications via legacy DNS exhibit relatively distinct patterns that can be used for network monitoring and anomaly detection purposes. The plaintext nature of DNS traffic makes the detection of C\&C-related activities reasonably achievable by way of traffic flow analysis and/or packet inspection \cite{JLeeWSNP2010}.

For example, features (\eg temporal activities, composition of domain name strings, DNS header flags) extracted from DNS traffic can be used to detect malware-infected devices. 
As pointed out by A. Udiyono \textit{et al.} \cite{AUdiyonoICONETSI2020}, botnet devices (as opposed to benign hosts) often send regular updates via periodic DNS messages on their status to the master servers, resulting in a high volume of DNS queries.
Also, domain names owned and operated by malicious actors are likely to be short-lived with low TTL values (\ie also known as disposable domains \cite{YZengCN2021}). That way, they can frequently change their DNS mapping and hence bypass static measures like blocklisting. In addition, devices that are infected by certain types of malware query for a relatively high number of distinct domain names due to the use of domain generation algorithms (DGA) that automatically emits a sophisticated set of domain names as the rendezvous points with C\&C servers. 
Using the DNS features discussed above, A. Udiyono \textit{et al.} \cite{AUdiyonoICONETSI2020} were able to detect botnet traffic with a relatively high accuracy ($\approx$90\%).

DNS features have proven effective in detecting malicious domain names used in C\&C communications. L. Bilge \textit{et al.} \cite{LBilgeNDSS2011} extracts 15 DNS traffic features of four categories, including time-based, DNS answer-based, TTL value-based, and domain name-based properties, which collectively detect malicious domain names used for C\&C. The authors developed machine learning models (trained on the identified features) with a large dataset containing over 100 billion DNS requests collected from an ISP network. The evaluation results show a close to perfect accuracy (more than 99.5\%) can be achieved in detecting malicious domains.

\begin{table*}[t!]
	\caption{Detecting C\&C communications: example DNS features and their availability for plaintext and encrypted DNS.}
	\centering
	\normalsize
	\renewcommand{\arraystretch}{1.2}
	\begin{tabular}{|p{9cm}|c|c|}
		\hline
		\rowcolor[rgb]{ .906,  .902,  .902} 	\textbf{Features}      & \textbf{Plaintext DNS} & \textbf{Encrypted DNS} \\ \hline
		Query packet size \cite{AUdiyonoICONETSI2020,CJDietrichEC2ND} &  \cellcolor[rgb]{ .502,  .996,  .502} Yes  \cmark     &  \cellcolor[rgb]{ .502,  .996,  .502}Yes   \cmark         \\ \hline
		Domain TTL value \cite{AUdiyonoICONETSI2020}  &  \cellcolor[rgb]{ .502,  .996,  .502}Yes  \cmark& \cellcolor[rgb]{ .996,  .502,  .502}No \xmark \\ \hline
		Number of distinct domain names \cite{AUdiyonoICONETSI2020,LBilgeNDSS2011} &  \cellcolor[rgb]{ .502,  .996,  .502}Yes  \cmark &\cellcolor[rgb]{ .996,  .502,  .502}No \xmark \\ \hline
		Number of client IP addresses \cite{LBilgeNDSS2011} &  \cellcolor[rgb]{ .502,  .996,  .502}Yes  \cmark&  \cellcolor[rgb]{ .502,  .996,  .502}Yes  \cmark\\ \hline
		Temporal and spatial query pattern \cite{LBilgeNDSS2011,CJDietrichEC2ND} &  \cellcolor[rgb]{ .502,  .996,  .502}Yes  \cmark&  \cellcolor[rgb]{ .502,  .996,  .502}Yes  \cmark\\ \hline
		Fraction of numerical characters in domain names \cite{LBilgeNDSS2011} &  \cellcolor[rgb]{ .502,  .996,  .502}Yes  \cmark&\cellcolor[rgb]{ .996,  .502,  .502}No \xmark\\ \hline
		Length of the longest meaningful substring (LMS) \cite{LBilgeNDSS2011} &  \cellcolor[rgb]{ .502,  .996,  .502}Yes  \cmark& \cellcolor[rgb]{ .996,  .502,  .502}No \xmark \\ \hline
		Entropy of domain names \cite{CJDietrichEC2ND} &  \cellcolor[rgb]{ .502,  .996,  .502}Yes  \cmark& \cellcolor[rgb]{ .996,  .502,  .502}No \xmark\\ \hline
		Volume of NXDOMAIN response \cite{YWangCN2021} &  \cellcolor[rgb]{ .502,  .996,  .502}Yes  \cmark& \cellcolor[rgb]{ .996,  .502,  .502}No \xmark\\ \hline
	\end{tabular}
	\label{tab:featureCC}
\end{table*}

In addition, by employing attributes of DNS traffic, researchers have identified a group of hosts managed certain malware families. 
C. J. Dietrich \textit{et al.} \cite{CJDietrichEC2ND} shows how traffic characteristics like the entropy of queried domain names or aggregate behavioral activity profiles (\eg DNS response rates) enabled the authors to detect C\&C via DNS and determine their malware families.

\subsubsection{C\&C Communications via Encrypted DNS}

Recent evidence from the security community \cite{CPatsakisCS2020} has shown that malware developers are increasingly exploiting encrypted DNS (\ie DoH) for C\&C communications, bypassing the existing tools for security monitoring. For example, in 2019, the security team of 360 Netlab reported the existence of Godlua backdoor malware \cite{Godlua2019} that uses DoH for C\&C communications. According to \cite{Godlua2019}, several vendors mark Godlua as mining-related trojan malware while also involved in DDoS attacks.
Malware strains that use encrypted DNS for C\&C communications continue to evolve and display more dynamic and stealthy behaviors to reduce their chance of being detected. 
As reported by the Proofprint Threat Insight Team \cite{PsiXBot2019}, the ``.NET-based'' malware PsiXBot that emerged to perform C\&C communications via the resolution of certain malicious domains now uses Google's public DoH resolver, and hence does not look suspicious anymore.
Therefore, network security appliances become ineffective when applying their pre-populated blocklists to those malicious C\&C-related queries.

\textbf{Implications of encryption for C\&C detection:} The trend of C\&C via encrypted DNS inevitably introduces significant challenges to defenders as they become unable to extract most of the key DNS features as in legacy plaintext DNS \cite{KBumanglagICICT2020}. 
Table~\ref{tab:featureCC} summarizes key features for detecting C\&C communications over plaintext DNS traffic (already discussed in \S\ref{sec:CCPlaintext}) and challenges (lack of their availability) when applied to encrypted DNS. 

Unsurprisingly, those features that require inspection of DNS packet headers and payloads (\eg query name, query type, error code, resource records, and domain TTL) are not available anymore due to content encryption. In contrast, obtaining features from unencrypted IP headers (\eg client IP addresses in Table~\ref{tab:featureCC}), aggregate network profiles (\eg query packet volume), and time-series patterns \cite{MMontazeriShatooriDASC2020} may still be feasible \cite{MGrillIM2015} for detecting C\&C over encrypted DNS.

\begin{figure*}[!t]
	\begin{center}
		\includegraphics[width=1\textwidth]{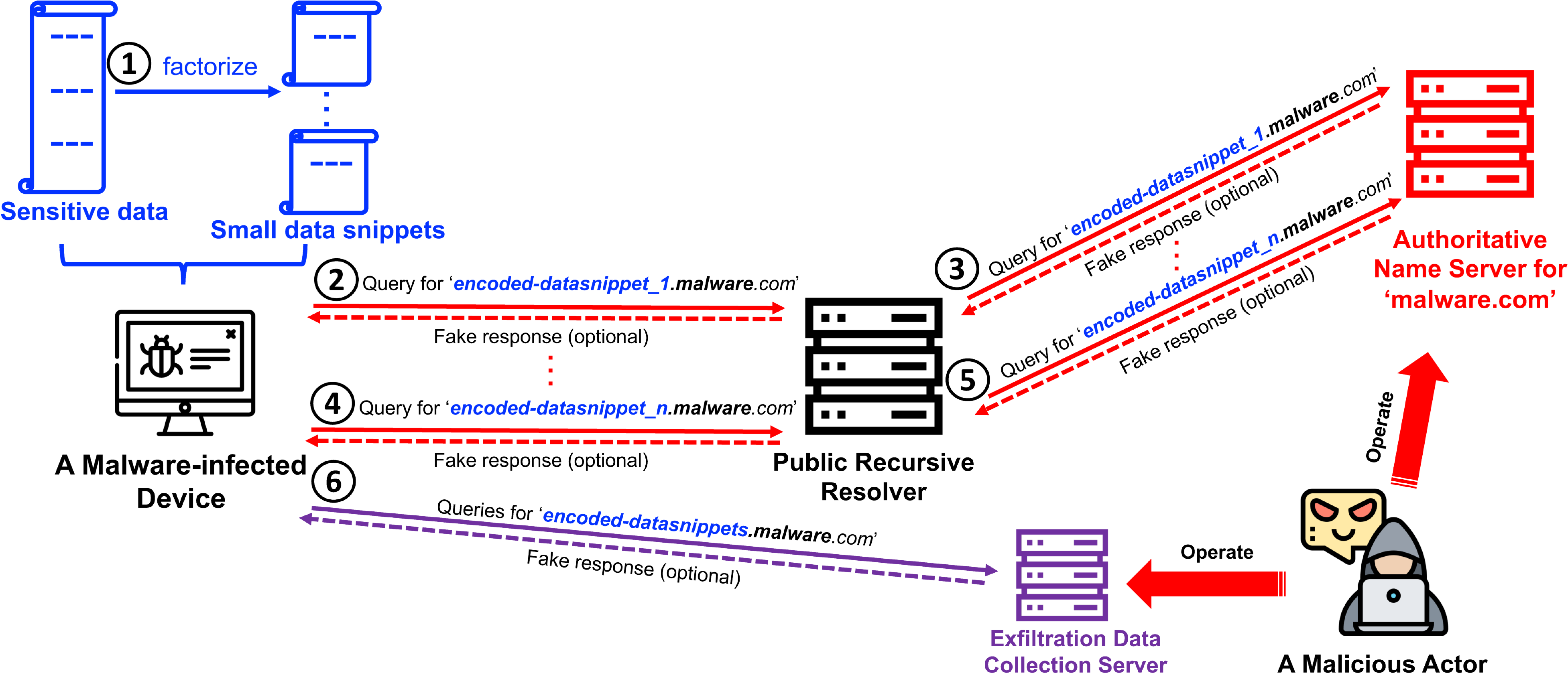}
		\caption{A visual example of data exfiltration via DNS from malware-infected devices.}
		\label{fig:DNSExfil}
	\end{center}
\end{figure*}

\subsection{Data Exfiltration (Tunneling)}\label{sec:Exfiltration}

Malware may retrieve sensitive and critical data (\eg user credentials, credit card details, business financial, or medical records) from the infected devices and send it to unauthorized external entities. This process is known as ``data exfiltration'' by the security community \cite{FUllahJNCA2018}.

\subsubsection{Data Exfiltration via Plaintext DNS}

Commercial security middleware and appliances, operation on the network of ISPs and/or enterprises, may be empowered to prevent or detect data thefts \cite{CISCODataExfiltration} by way of traffic inspection.
However, DNS communication is relatively poorly policed by organizations (compared to services like email, FTP, and HTTP) and has been exploited by cyber-criminals to maintain covert communication channels with compromised hosts. Such stealthy methods, leveraging the DNS protocol, is known as DNS exfiltration or tunneling \cite{ANadlerCS2019} .

\textbf{DNS exfiltration:}
Fig.~\ref{fig:DNSExfil} illustrates a typical process of DNS exfiltration,  with a sequence of events annotated by circled numbers. For the first step (the top-left region of Fig.~\ref{fig:DNSExfil}), a malware-infected device factorizes the sensitive data into small snippets that are suitable to be encoded into (the subdomain of) DNS query names. 
From step \ciao{2} to \ciao{5}, the device crafts and sends a series of DNS queries with their subdomain names carrying the encoded data (\ie {\myverb{encoded-datasnippet\_1}} till {\myverb{encoded-datasnippet\_n}}) belonging to a malicious domain (\ie {\myverb{malware.com}}) managed by an external attacker. These queries reach the malicious name server ({\myverb{malware.com}}) through private/public recursive resolvers and might get dummy responses as the acknowledgments of received data. Exfiltrating data via DNS recursive resolvers makes the delivery path seemingly legitimate, however, recursive resolvers may block some of the reported primary domains to stop such misuses.
We note that malicious actors may choose to directly send exfiltration DNS queries to a collection server managed by them (step \ciao{6} in Fig.~\ref{fig:DNSExfil}).
Finally, upon arrival of all malicious DNS queries (sourced from the infected device), the attacker can reconstruct the exfiltrated data file by decoding and combining all snippets extracted from the subdomain names.

\begin{table*}[!t]
	\caption{Examples of a short benign domain name, a long benign domain name, and a data exfiltration domain name \cite{JAhmedTNSM2020}.}
	\centering
	\normalsize
	\renewcommand{\arraystretch}{1.2}
	\begin{tabular}{|p{1.1cm}|p{13.2cm}|}
		\hline
		\rowcolor[rgb]{ .906,  .902,  .902} 	\textbf{Type}              & \textbf{Domain name} \\ \hline
		Benign           &  \cellcolor[rgb]{ .502,  .996,  .502}{\myverb{www.google.com}} \\ \hline
		Benign           &   \cellcolor[rgb]{ .54,  0.81, .94}{\myverb{p4-ces3lawazdkbw-qlrq5qalxdt7ytcq-385202-i1-v6exp3.ds.metric.gstatic.com}}\\ \hline
		Exfil.& \cellcolor[rgb]{ .996,  .502,  .502}{\myverb{PzMnPios0D3nOCwu0zomPS4nNjovPS8uOzsnNCstODkjOcwoMwAA.29a.de}} \\ \hline
	\end{tabular}
	\label{fig:ExfilNormalDomainName}
\end{table*}

\textbf{Countermeasures against data exfiltration via plaintext DNS:}
Research works on detection of data exfiltration over plaintext DNS have proven to be successful. They generally employed statistical features extracted from query names to train machine learning classifiers \cite{BSabirCS2022} to distinguish benign from malicious queries. 

Let us have a look at some of these queries. Table~\ref{fig:ExfilNormalDomainName} shows three sample query names (benign and exfiltration), as reported by by J. Ahmed \textit{et al.} \cite{JAhmedTNSM2020}, illustrating the difference between benign and malicious query names in the context of plaintext DNS. We note that such distinction may not be evident with encrypted DNS queries.
It can be seen the exfiltrated query name (the thrid row highlighted in red) contains a fairly long subdomain string with 52 English letters, whereas a typical benign subdomain only contains several English letters (\eg three for {\myverb{www}} in the first row). Note that some benign domain names may also have a long subdomain string (the second row highlighted in blue).
However, exfiltration strings generated by encoding algorithms look random with a mix of numbers and uppercase/lowercase letters. In contrast, the long benign subdomains often display more meaningful patterns indicative of their service categories/indexes (the first row highlighted in green).

Based on large-scale data analysis of real queries for both benign and suspicious/malicious domains, J. Ahmed \textit{et al.} \cite{JAhmedTNSM2020} identified eight features of individual domain names to detect anomalous string patterns in exfiltration DNS queries. They used features like the count of uppercase characters and characters in subdomain names. The authors trained a one-class classifier on benign data that achieved an overall detection accuracy of more than 98\%.

Using a different approach, work in \cite{BSabirCS2022} focused on other aspects of the DNS traffic, considering features like spatial and temporal statistics of query packets from a given host to detecting DNS exfiltration from malware-infected hosts.

\subsubsection{Data Exfiltration via Encrypted DNS}
The deployment of DNS encryption techniques and their supporting infrastructures (\eg encryption-enabled public resolvers) provide a solid basis for data exfiltration over encrypted DNS protocols.
Therefore, it is not surprising to witness their occurrence, such as the one described in \cite{CCimpanu2020}. Oilrig (or APT34), a hacking group, was reported in 2020 as the first known malicious actor that managed to perform data exfiltration over DoH by advancing an open-source exfiltration toolkit \cite{ExfilDoHTool2}. The victim of that DoH-based data exfiltration may include a pharmacy company when it announced its commencement of research for the treatment of COVID-19, according to \cite{CCimpanu2020}.
Many open-source tools \cite{ExfilDoHTool} exist for data exfiltration over DoH on the Internet and hence becomes relatively easy for malicious actors. Therefore, enterprises that host critical data are alarmed \cite{CISCODataExfiltration} to stay vigilant for emerging threats before getting harmed.

\begin{table*}[t!]
	\caption{Detecting data exfiltration: representative DNS features extracted from domain names and their availability in plaintext and encrypted DNS traffic.}
	\centering
	\normalsize
	\renewcommand{\arraystretch}{1.2}
	\begin{tabular}{|p{9.5cm}|c|c|}
		\hline
		\rowcolor[rgb]{ .906,  .902,  .902} 	\textbf{Features}       & \textbf{Plaintext DNS} & \textbf{Encrypted DNS} \\ \hline
		Total count of characters in fully qualified domain names (FQDN)  \cite{DHaddonICGS32019,JAhmedTNSM2020}  &  \cellcolor[rgb]{ .502,  .996,  .502}Yes   \cmark         &\cellcolor[rgb]{ .996,  .502,  .502}No \xmark\\ \hline
		Count of characters in sub-domain names \cite{JAhmedTNSM2020} & \cellcolor[rgb]{ .502,  .996,  .502}Yes       \cmark     &\cellcolor[rgb]{ .996,  .502,  .502}No \xmark\\ \hline
		Count of uppercase characters in domain names \cite{JAhmedTNSM2020} & \cellcolor[rgb]{ .502,  .996,  .502}Yes  \cmark          &\cellcolor[rgb]{ .996,  .502,  .502}No \xmark\\ \hline
		Count of numerical characters in domain names \cite{JAhmedTNSM2020} &  \cellcolor[rgb]{ .502,  .996,  .502}Yes  \cmark&\cellcolor[rgb]{ .996,  .502,  .502}No \xmark\\ \hline
		Entropy of domain names \cite{DHaddonICGS32019,JAhmedTNSM2020} &  \cellcolor[rgb]{ .502,  .996,  .502}Yes  \cmark&\cellcolor[rgb]{ .996,  .502,  .502}No \xmark\\ \hline
		Number of domain name labels \cite{JAhmedTNSM2020} & \cellcolor[rgb]{ .502,  .996,  .502}Yes  \cmark&\cellcolor[rgb]{ .996,  .502,  .502}No \xmark\\ \hline
		Maximum domain name labels \cite{JAhmedTNSM2020} & \cellcolor[rgb]{ .502,  .996,  .502}Yes  \cmark& \cellcolor[rgb]{ .996,  .502,  .502}No \xmark\\ \hline
		Average domain name labels \cite{JAhmedTNSM2020} & \cellcolor[rgb]{ .502,  .996,  .502}Yes \cmark & \cellcolor[rgb]{ .996,  .502,  .502}No \xmark \\ \hline
		
	\end{tabular}
	\label{tab:featureExfil}
\end{table*}

\textbf{How DNS encryption impacts data exfiltration detection:} 
The use of encrypted DNS in data exfiltration attacks inevitably makes detecting them more difficult than those leveraging plaintext DNS. Methods for detecting C\&C over encrypted DNS may still be able to resort to patterns in signaling packets (\ie malware-infected devices are likely to connect with their C\&C servers periodically \cite{XHuDSN2016}). 
However, existing methods for detecting DNS data exfiltration that primarily analyze the string patterns in query names become ineffective after encryption. Table~\ref{tab:featureExfil} summarizes features\footnote{The entropy of domain names has already been discussed in \S\ref{sec:CC} for C\&C detection.} extracted from query names \cite{JAhmedTNSM2020,DHaddonICGS32019} that can recognize (with decent confidence) the DNS exfiltration over plaintext DNS. It can be seen in the last column of this table that all required features become unavailable after DNS encryption.

\section{Inference Techniques for Encrypted DNS Traffic}\label{sec:SurveyDetection}
As discussed in \S\ref{sec:SurveyMalware}, malware leverages encrypted DNS protocol to bypass inspection-based detection and mitigation techniques used for C\&C communications and data exfiltration.
Although the current countermeasures against such misuses have not matured, some promising attempts to analyze encrypted DNS traffic for classification (not necessary for security purposes) provide useful lessons and references.
In this section, we start by elaborating on the current research works in detecting encrypted DNS packets from HTTPS traffic streams and classifying their types (\eg benign or malicious) in \S\ref{sec:DetectingEDNS}. Next, we review the developed methods in fingerprinting user profiles by analyzing encrypted DNS traffic, which can be a useful reference in developing methods to identify malware-infected hosts (\S\ref{sec:FingerprintingViaEDNS}).

\subsection{Detecting and Classifying Encrypted DNS Traffic}\label{sec:DetectingEDNS}
Plaintext DNS, DoT, and DoQ have their dedicated service port numbers as {\myverb{UDP/53}}, {\myverb{TCP/853}}, and {\myverb{UDP/784\&8853}}, respectively, which can be easily detected by inspecting headers. 
However, DoH encapsulates its DNS content in HTTPS packets, and thus it shares the same service port {\myverb{TCP/443}} with other HTTPS-based applications (\eg web browsing). Unsurprisingly, the detection of DoH packets from HTTPS traffic is a nontrivial problem. 

One may label an HTTPS flow as DoH if its source or destination IP address is one of the well-known DoH recursive resolvers. However, such methods inevitably: (a) mislabel non-DoH HTTPS flows towards the IP address of popular resolvers (\eg Google's DoH resolver shares the IP address {\myverb{8.8.8.8}} with an HTTPS-based website \cite{GoogleDNS} for manual DNS lookups via a web interface), and (b) will miss DoH flows to unpopular servers.
As pointed out in \cite{CPatsakisCS2020}, the analysis of traffic patterns seems to be a promising approach to detect and classify DoH packets due to some of their distinct characteristics. In what follows, we discuss the related research efforts in more detail.

\subsubsection{Detecting DoH from HTTPS Traffic}
D. Vekshin \textit{et al.} \cite{DVekshinARES2020} pointed out that accurate recognition of DoH is a precursor step for detecting encrypted DNS-based malware activities.
They explored the possibility of using machine learning techniques to differentiate DoH from HTTPS traffic. 
To this end, the authors analyzed the identifiable flow characteristics of DoH.
First, clients often establish a single DoH connection to their recursive resolver and retain it for a number of domain lookups. Therefore, the flow duration of DoH is likely to be longer than that of typical HTTPS connections (\eg web browsing). Although video streaming and large file downloading via HTTPS may also have long durations, their volume and data rate are expected to be higher than DoH flows. Second, the variance of DoH packet sizes is much lower than other HTTPS applications. Third, DoH packets often come in bursts during a connection depending on users' interactions, but the total number of DNS packets (and the corresponding volume) in each burst is not large.
Finally, unlike other applications, a DoH flow often has quite similar (or symmetric) volume in both directions.

\begin{table*}[!t]
	\caption{Key flow features (summarized from information in \cite{DVekshinARES2020}) and their expected value range for DNS over HTTPS (DoH), web browsing, file downloading, and video streaming.}
	\centering
	\normalsize
	\renewcommand{\arraystretch}{1.2}
	\begin{tabular}{|p{5cm}|p{2cm}|p{2cm}|p{2cm}|p{2cm}|}
		\hline
		\rowcolor[rgb]{ .906,  .902,  .902} 	\textbf{Features}              & \textbf{DNS over HTTPS} & \textbf{Web browsing} & \textbf{File downloading} & \textbf{Video streaming} \\ \hline
		Flow duration         &\cellcolor[rgb]{ .996,  .502,  .502}Long & \cellcolor[rgb]{ .502,  .996,  .502}Short &\cellcolor[rgb]{ .996,  .502,  .502}Long & \cellcolor[rgb]{ .996,  .502,  .502}Long\\ \hline
		Flow volume            & \cellcolor[rgb]{ .502,  .996,  .502}Small & \cellcolor[rgb]{ .996,  .996,  .0}Medium  & \cellcolor[rgb]{ .996,  .502,  .502}Large & \cellcolor[rgb]{ .996,  .502,  .502}Large\\ \hline
		Variance of packet sizes &\cellcolor[rgb]{ .502,  .996,  .502}Small & \cellcolor[rgb]{ .996,  .502,  .502}Large & \cellcolor[rgb]{ .996,  .996,  .0}Medium & \cellcolor[rgb]{ .996,  .502,  .502}Large\\ \hline
		Ratio of burst (active) and pause (idle) & \cellcolor[rgb]{ .996,  .502,  .502}High & \cellcolor[rgb]{ .502,  .996,  .502}Small & \cellcolor[rgb]{ .502,  .996,  .502}Small & \cellcolor[rgb]{ .996,  .502,  .502}High \\ \hline
		Number of packets in each burst & \cellcolor[rgb]{ .996,  .996,  .0}Medium& \cellcolor[rgb]{ .996,  .502,  .502}High & \cellcolor[rgb]{ .996,  .502,  .502}High & \cellcolor[rgb]{ .996,  .502,  .502}High \\ \hline
		Symmetry of requests and responses & \cellcolor[rgb]{ .996,  .502,  .502}High & \cellcolor[rgb]{ .502,  .996,  .502}Low & \cellcolor[rgb]{ .502,  .996,  .502}Low & \cellcolor[rgb]{ .502,  .996,  .502}Low \\ \hline
	\end{tabular}\label{tab:DoHvsHTTPS}
\end{table*}

Observing some of unique characteristics of DoH flows (mentioned above) as well as their temporal behaviors, D. Vekshin \textit{et al.} \cite{DVekshinARES2020} identified 19 features capturing identifiable patterns in DoH flows.
We summarize and aggregate in Table~\ref{tab:DoHvsHTTPS} those key features and show their expected value range for DoH and three typical HTTPS-based applications, including web browsing, file downloading, and video conferencing.
With those key flow-level attributes, the authors developed machine learning classifiers to distinguish DoH from generic HTTPS traffic that yields a very high accuracy of more than 99\% on their dataset publicly available at \cite{vekshin_dmitrii_2020_3906526}.
Similarly, L. Csikor \textit{et al.} \cite{LCsikorESP2021} built their machine learning models using features of packet sizes and inter-arrival times that can differentiate DoH from website browsing with more than 90\% accuracy.

\subsubsection{Classifying Benign and Malicious DoH Traffic}
With the rising use of DoH in malicious activities such as C\&C and data exfiltration \& tunneling, classifying benign and malicious DoH traffic becomes increasingly essential.

C. Patsakis \textit{et al.} \cite{CPatsakisCS2020} performed time-series modeling via Hodrick-Prescott (HP) filter on DoT and DoH response sizes from benign and malware-infected hosts.
They observed distinct patterns in time-series signals of packet size to identify hosts performing C\&C communications and classify their malware families.

C. Kwan \textit{et al.} \cite{CKwan2021FOCI} developed a threshold-based method on packet size, packet rate, and throughput to differentiate DoH tunneling from benign DoH communications. Their method achieved a perfect 100\% accuracy in detecting DoH tunneling generated by a popular tool called ``dnstt'' \cite{DNSTTTool}. The authors demonstrated that DNS tunneling could only bypass their detection method if its rate is significantly reduced by a factor of 27.

M. MontazeriShatoori \textit{et al.} \cite{MMontazeriShatooriDASC2020} detects DoH tunneling by combining time-series (\ie a sequence of packet size, count, duration, and inter-arrival time), header (\ie unencrypted handshake information), and statistical features (\eg mean, median, and variance of packet sizes) of HTTPS traffic.
The authors make their dataset ``CIRA-CIC-DoHBrw-2020''  publicly available \cite{CIRACICDoHBrw2020}, containing more than 100M packets of generic HTTPS, benign DoH, and malicious DoH. Their traffic traces are generated by two browsers (Chrome and Firefox), three DoH exfiltration tools (Iodine, DNS2TCP, and DNScat2), and four public DoH resolvers (Adguard, Cloudflare, Google, and Quad9). The authors developed machine-learning models with a hybrid set of features from a rich and diversified dataset that achieved over 99\% precision in malicious DoH traffic detection.
Using the ``CIRA-CIC-DoHBrw-2020'' dataset, three follow-up research works \cite{SKSingh3ICT2020,YMBanadakiJCSA2020,MBehnkeAccess2021} demonstrated the effectiveness of various machine learning algorithms such as Logistic Regression (LR), Random Forest (RF), K-Nearest Neighbors (KNN),  XGBoost, Light gradient boosting machine (LGBM) using 34 traffic features originally identified by M. MontazeriShatoori \textit{et al.} \cite{MMontazeriShatooriDASC2020}.

\subsection{Profiling User Activity by Analyzing Encrypted DNS Traffic}\label{sec:FingerprintingViaEDNS}

In addition to the detection of encrypted DNS traffic from generic HTTPS traffic, some emerging research works analyze encrypted DNS traffic to profile the behavior of hosts (\ie users) -- insights such as websites they visit and operating systems they use. To the best of our knowledge, there is no such work for detecting malware-infected hosts. That said, current research arts provide valuable references and insights for solving cybersecurity problems. Therefore, we discuss them as follows. 

R. Houser \textit{et al.} in \cite{RHouserCoNEXT2019} fingerprints the history of visited websites by hosts using temporal patterns of DoT packet sequences.
The authors identify important features that fall into nine categories, including length of query and response, number of queries and responses, volume of queries and responses, time intervals, total transmission time, sequence of DNS packets of uninterrupted queries and responses, number of DNS messages in each TLS record, query rate, and time to receive the first \textit{N} bytes from the resolver. By selecting and tuning optimal models, the authors are able to achieve decent performance, false negative and false positive rates of less than 17\% and 0.5\%, respectively, in identifying the visited website of a host. Moreover, they showed how their techniques work for padded DoT messages \cite{rfcDNSPadding2018} with around 99\% true positive rate and 42\% true negative rates.

Similarly, S. Siby \textit{et al.} \cite{SDSibyNDSS2020,siby2018dns} proposed a machine learning-based method to fingerprint website visits of a host by analyzing DoH traffic. For a DoH flow, they use n-grams (\ie a contiguous sequence of n queries and responses) of TLS record lengths and burst-lengths (\ie the total length of consecutive packets in the same direction) as key features to train a random forest classifier. Evaluation results showed that their approach achieved more than 90\% accuracy in fingerprinting website visits. The authors demonstrated that client locations, different DoH resolvers, client operating systems, and browser applications only have a minor impact on the classification performance. However, the use of perfect padding (currently not widely adopted by the stakeholders) deteriorates their model's prediction, giving a poor accuracy of less than 10\%.

\begin{table*}[!t]
	\caption{Key features and their applicability \& effectiveness (reported in \cite{RHouserCoNEXT2019,SDSibyNDSS2020,MMuhlhauserARES2021}) for user profiling when extracted from padded/unpadded DoT/DoH traffic.}
	\centering
	\normalsize
	\renewcommand{\arraystretch}{1.2}
	\begin{tabular}{|p{5.7cm}|c|c|c|c|}
		\hline
		\rowcolor[rgb]{ .906,  .902,  .902} 	\textbf{Features}              & \textbf{Unpadded DoH} &\textbf{Unpadded DoT} & \textbf{Padded DoH} & \textbf{Padded DoT}\\ \hline
		Query and response length   \cite{RHouserCoNEXT2019}      &\cellcolor[rgb]{ .502,  .996,  .502}Applicable & \cellcolor[rgb]{ .502,  .996,  .502}Applicable&\cellcolor[rgb]{ .996,  .996,  .0}Not effective &\cellcolor[rgb]{ .996,  .996,  .0}Not effective \\ \hline
		Number of queries and responses  \cite{RHouserCoNEXT2019}      &\cellcolor[rgb]{ .502,  .996,  .502}Applicable & \cellcolor[rgb]{ .502,  .996,  .502}Applicable&\cellcolor[rgb]{ .502,  .996,  .502}Applicable & \cellcolor[rgb]{ .502,  .996,  .502}Applicable \\ \hline
		Inter-arrival time   \cite{RHouserCoNEXT2019}    &\cellcolor[rgb]{ .502,  .996,  .502}Applicable& \cellcolor[rgb]{ .502,  .996,  .502}Applicable&\cellcolor[rgb]{ .502,  .996,  .502}Applicable &\cellcolor[rgb]{ .502,  .996,  .502}Applicable \\ \hline
		Transmission time  \cite{RHouserCoNEXT2019}     &\cellcolor[rgb]{ .502,  .996,  .502}Applicable&\cellcolor[rgb]{ .502,  .996,  .502}Applicable &\cellcolor[rgb]{ .502,  .996,  .502}Applicable & \cellcolor[rgb]{ .502,  .996,  .502}Applicable \\ \hline
		Order of query and response packets   \cite{RHouserCoNEXT2019}     &\cellcolor[rgb]{ .502,  .996,  .502}Applicable & \cellcolor[rgb]{ .502,  .996,  .502}Applicable&\cellcolor[rgb]{ .502,  .996,  .502}Applicable &\cellcolor[rgb]{ .502,  .996,  .502} Applicable \\ \hline
		Volume of queries and responses \cite{RHouserCoNEXT2019} &\cellcolor[rgb]{ .502,  .996,  .502}Applicable& \cellcolor[rgb]{ .502,  .996,  .502}Applicable&\cellcolor[rgb]{ .996,  .996,  .0} Not effective &\cellcolor[rgb]{ .996,  .996,  .0} Not effective \\ \hline
		Number of DNS message in TLS records  \cite{RHouserCoNEXT2019}      &\cellcolor[rgb]{ .996,  .502,  .502}Not applicable & \cellcolor[rgb]{ .502,  .996,  .502}Applicable &\cellcolor[rgb]{ .996,  .502,  .502}Not applicable &\cellcolor[rgb]{ .502,  .996,  .502}Applicable \\ \hline
		Query rate \cite{RHouserCoNEXT2019}       &\cellcolor[rgb]{ .502,  .996,  .502}Applicable& \cellcolor[rgb]{ .502,  .996,  .502}Applicable& \cellcolor[rgb]{ .502,  .996,  .502}Applicable &\cellcolor[rgb]{ .502,  .996,  .502}Applicable \\ \hline
		Time to receive the first \textit{N} bytes \cite{RHouserCoNEXT2019}       &\cellcolor[rgb]{ .996,  .502,  .502}Not applicable &\cellcolor[rgb]{ .502,  .996,  .502}Applicable &\cellcolor[rgb]{ .996,  .502,  .502}Not applicable &\cellcolor[rgb]{ .502,  .996,  .502}Applicable \\ \hline
		N-grams of TLS record lengths \cite{SDSibyNDSS2020}&\cellcolor[rgb]{ .502,  .996,  .502}Applicable & \cellcolor[rgb]{ .996,  .996,  .0}Not effective  &\cellcolor[rgb]{ .996,  .996,  .0}Not effective & \cellcolor[rgb]{ .996,  .996,  .0}Not effective \\ \hline
		N-grams of burst sizes \cite{SDSibyNDSS2020}&\cellcolor[rgb]{ .502,  .996,  .502}Applicable & \cellcolor[rgb]{ .996,  .996,  .0}Not effective &\cellcolor[rgb]{ .996,  .996,  .0}Not effective & \cellcolor[rgb]{ .996,  .996,  .0}Not effective \\ \hline
		N-grams of DNS sequences \cite{MMuhlhauserARES2021}&\cellcolor[rgb]{ .502,  .996,  .502}Applicable &\cellcolor[rgb]{ .502,  .996,  .502}Applicable &\cellcolor[rgb]{ .996,  .996,  .0}Not effective  & \cellcolor[rgb]{ .996,  .996,  .0}Not effective \\ \hline
	\end{tabular}\label{tab:FingerprintFeature}
\end{table*}

Apart from website fingerprinting, Segram, as presented by M. Muhlhauser \textit{et al.} \cite{MMuhlhauserARES2021}, is able to identify Android applications for smartphones and IoT devices used by clients through the analysis of their DoT or DoH traffic.
Apart from website fingerprinting, Segram \cite{MMuhlhauserARES2021} is able to identify Android applications for smartphones and IoT devices used by clients through the analysis of their DoT or DoH traffic. The authors used the n-grams of DNS sequences (message sizes and interarrival time) as primary features of their classification model, which outperforms their counterparts \cite{SDSibyNDSS2020} employing other features such as n-grams of TLS record sizes and burst lengths. More specifically, their developed model achieved more than 90\% accuracy on the traffic traces (publicly available at \cite{MMulhauser2021}) of 118 Android apps from ten DoT/DoH resolvers. They highlighted that their method is proven to be relatively effective when applied to padded encrypted DNS, giving over 72\% classification accuracy.

In addition to profiling the user behaviors by analyzing encrypted DNS traffic on the network, G. Varshney \textit{et al.} \cite{GVarshneyCOMSNETS2021} showed that how they can obtain DoH lookups before they get encrypted by passively monitoring the RAM usage on the client devices. 
It is important to note that DNS encryption promises to protect data privacy in transit, but organizations and/or users cannot expect it to prevent such strong attacks when a user device is infected/compromised.

In Table~\ref{tab:FingerprintFeature}, we summarize the applicability and efficacy of key features of unpadded DoH, unpadded DoT, padded DoH, and padded DoT traffic in order to profile users online activity as reported by the above research works.  For a certain traffic type, a feature is labeled as ``applicable'' if it can be extracted; otherwise, it is labeled as ``not applicable''.  In addition, a feature that can be computed but leads to relatively poor classification performance (\ie accuracy less than 80\% as reported by the respective literature) is marked as ``not effective'' in Table~\ref{tab:FingerprintFeature}. 
For example, as stated by M. Muhlhauser \textit{et al.} \cite{MMuhlhauserARES2021}, classifications using n-grams of DNS sequences can only achieve 72\% accuracy for padded DoH and DoT; thus, they are labeled as ``not effective'' in the respective cells.
We believe that those important characteristics of encrypted DNS traffic are still valuable in future user profiling works for cybersecurity use-cases like detecting malware-infected hosts.

\section{Discussion and Future Directions}\label{sec:discussion}
With the understanding of current development and issues of DNS encryption and some of the risks of malware misuse, we now discuss potential research directions worthwhile to be explored in future works.

\subsection{Performance Enhancements}
Although DNS encryption promises privacy and security benefits to users and service providers, overheads of TCP acknowledgments and TCP/SSL handshakes in DoT and DoH can led to some performance degradation (\eg higher query resolution times are observed in the distribution of real-world measurements discussed in \S\ref{sec:performanceAnalysis}, such as by A. Hounsel \textit{et al.} \cite{AHounselANRW2021}) compared with plaintext DNS over UDP. 
Such performance degradation is more evident to clients under unstable network conditions in rural areas and/or wireless networks. Thus, it becomes one of the key reasons for the low adoption of DNS encryption.
DoQ, which is currently in the draft stage, seems to be a faster DNS encryption protocol due to its zero-RTT feature.
Apart from the protocol-level optimization, future research for performance enhancement mechanisms on client applications (\eg browsers), network equipment (\eg routers), and resolvers is necessary to boost the public adoption of DNS encryption.

\subsection{Managing Security and Privacy Vulnerability}
As discussed in \S\ref{sec:secIssues} , there are still privacy issues that remain unresolved or are newly introduced by DNS encryption. We now recap/highlight those issues and outline some potential directions of research to address them.

First, although DNS communications between clients and resolvers are secured against third-party eavesdropping, a compromised (or malicious) resolver could still put user privacy at risk. Furthermore, current DoT/DoH/DoQ resolvers are dominated by a limited number of servers owned and operated by major service providers. Such centralization could lead to data monopolization that ultimately harms clients' privacy. Therefore, developing privacy-preserving methods against data leakage from the resolver is a valuable direction -- ODNS \cite{PSchmittANRW2019} that detaches the proxy function from resolvers is an example of such methods.

Second, compared with plaintext DNS resolvers, DoH and DoT resolvers become vulnerable to a broader range of attacks such as TCP and TLS-based DDoS attacks in addition to typical DNS query floods given their need for establishing TLS/TCP connections with clients (legitimate or malicious). In addition, DNS encryption is not a silver bullet to prevent all types of attacks over DNS, such as injection attack \cite{PJeitnerSEC2021}. Undoubtedly, effectively securing the key infrastructure (\ie encrypted DNS resolvers) from potential DDoS attacks requires more sophisticated protection, mostly related to the management of TCP and TLS connections, than plaintext DNS.

Third, misconfiguring encryption settings on either client (\eg choice of encryption option) or resolver (\eg invalid SSL certificate or insecure TLS version) make the privacy protection promised by DNS encryption less effective. Therefore, a systematic way of sanity checking and/or verification on DNS encryption settings for user applications (\eg browsers and mobile applications) and resolver can be an avenue for research to explore. 

Last, third-party observers (or eavesdroppers) may still identify user profiles (\eg visited websites and used applications) by analyzing the traffic patterns of their encrypted DNS communications. Malicious usage of such techniques may result in leaking clients' private information. Some potential countermeasures such as privacy-preserving query scheduling techniques (\eg \cite{OAranaCN2021}) that obfuscate DNS lookup patterns of various applications are worthwhile to be further developed, protecting user privacy.

\subsection{Understanding Malicious Encrypted DNS Traffic}
As discussed in \S\ref{sec:SurveyMalware}, encrypted DNS has been misused for malware C\&C communications and data exfiltration to evade security appliances and middleware that perform payload inspection, looking for malicious signatures. With the increasing growth of DNS encryption, its misuse by malicious actors will likely become more frequent and diversified. Hence, there is a need for advanced methods to characterize malicious encrypted DNS communications, addressing emerging threats.

\subsection{Host Profiling Techniques for Malware Detection}
As discussed in \S\ref{sec:FingerprintingViaEDNS}, current host profiling techniques that analyze encrypted DNS traffic mainly focus on identifying visited websites and mobile applications. Existing works have shown that hosts display distinguishable patterns of encrypted DNS traffic when they perform certain activities (\eg visiting different websites).
They extract a comprehensive set of features, achieving acceptable accuracy in their respective tasks. 
We believe such profiling techniques can be employed to detect malware-infected hosts. For example, quantifying deviations from their baseline benign profile (anomaly detection) or employing multi-class classifications to identify specific types of malware misuses (\eg classes of malware or types of misuse) could be considered.

We note that the features (summarized in Table~\ref{tab:FingerprintFeature}), proven to be effective for other tasks, may also be useful in detecting malware activities via encrypted DNS.

\section{Related Surveys on DNS Security}\label{sec:relatedWork}
DNS security has been a popular topic, with many published surveys, each focusing on a specific topic ranging from the categorization of DNS attacks to methods for detecting DNS attacks or identifying certain types of DNS attacks. We discuss (in chronological order) some of the contemporary ones published after 2016.

Y. Zhauniarovich \textit{et al.} \cite{YZhauniarovichCS2018} reviewed data-driven techniques for detecting malicious domains with a focus on aspects of dataset collection (\ie the sources of DNS data, data enrichment, and ground truth), algorithm design (\ie feature engineering, detection mechanisms, and outcomes), and evaluation methodologies (\ie performance metrics and strategies).
S. Torabi \textit{et al.} \cite{STorabiCST2018} surveyed passive systems of DNS traffic analysis for detecting various DNS attacks across the Internet, including DNS protocol attacks (\eg DNS spoofing), DNS server attacks (\eg DDoS), and DNS abuse (\eg C\&C communications). The authors looked at various system designs and compared their objectives, scopes, detection approaches, analyzed traffic features, dataset, and evaluation methods. The survey also highlighted the practicality of the studied systems by considering their real-time performance.
N. U. Aijaz \textit{et al.} \cite{NUAijazDSS2021} studied DNS vulnerabilities, including cache poisoning attack, DDoS attack, spoofing, phishing, identity theft, forgery, eavesdropping, packet sniffing, tampering, and man-in-the-middle attack.
In their discussions, they also included the protocol-level security enhancements introduced for DNS (\ie DNSSEC, SSL certificates, and extended validation certificate).
Y. Wang \textit{et al.} \cite{YWangCN2021} surveyed (plaintext) DNS tunneling detection methods developed during years between 2006 and 2020. They highlight the statistical features (\ie payload-based and traffic-based) and detection mechanisms (\ie rule-based and model-based) used in each DNS tunnel detection method.

Our survey focuses on DNS encryption and highlights opportunities (\ie protecting privacy) and risks (\eg malware misuse) associated with encrypted DNS traffic. To the best of our knowledge, there is no survey on the topic of DNS encryption to date. We found all existing relevant survey papers only cover the landscape of plaintext DNS.

\section{Conclusion}\label{sec:conclusion}
This paper conducted a systematic and comprehensive review of academic research papers and industrial reports on the development and current status of DNS encryption, with a specific focus on its misuse by malware and potential countermeasure techniques.
We outlined the development of three DNS encryption protocols (including DoT, DoH, and DoQ), their current state of public adoption, and their performance. We discussed the security benefits and risks of DNS encryption techniques highlighted by the current literature.
Among security risks, we particularly focused on the potential malware misuses of encrypted DNS protocols, including C\&C and data exfiltration. Although detecting malicious activities over plaintext DNS has proven relatively successful by prior works using inspection DNS payloads, their applicability to encrypted traffic is yet to be determined.
Next, we studied existing works on the analysis of encrypted DNS traffic that can detect malware activities over encrypted DNS. The works are classified as either detecting encryption DNS packets from generic HTTPS traffic and classifying their types (\eg malicious or benign), or fingerprinting user profiles by analyzing encrypted DNS traffic of hosts. Those works provide solid starting points and valuable references for malware detection potentially applicable to encrypted DNS.
Inspired by prior works in the literature, we identified directions for future works, including performance enhancement, managing security and privacy issues, understanding the misuse of encrypted DNS by malware, and developing host profiling techniques to detect malware activities over encrypted DNS protocols.

\bibliographystyle{ACM-Reference-Format}
\bibliography{ReferencesSurvey}

\end{document}